\documentclass{aa}  
\usepackage[flushleft]{threeparttable}
\usepackage{xcolor}
\usepackage{commath}
\usepackage{CJKutf8}
\usepackage{url}
\usepackage{tikz}
\usepackage{amsmath}
\usepackage{amssymb}
\usepackage{graphicx}
\usepackage{txfonts}
\usepackage{bm}
\newcommand{\mi}{M_\mathrm{1,i}}
\newcommand{\qi}{q_\mathrm{i}}
\newcommand{\logpi}{\mathrm{log}\,P_\mathrm{orb,i}}
\newcommand{\logp}{\,\mathrm{log}\,P_\mathrm{orb}}
\def\porb{\,P_\mathrm{orb}}
\def\kms{\,\mathrm{km\,s}^{-1}}
\def\dd{\mathrm{ d\,}}
\def\mso{\,M_\odot}
\def\rso{\,R_\odot}
\def\lso{\,L_\odot}

\def\msoy{\, \mso~\mathrm{ yr}^{-1}}

\newcommand*{\vv}[1]{{\overrightarrow{#1}}}
\begin{document}

   \title{Evolution of wide O star binaries through their LBV stage
}
   
   \subtitle{Population synthesis with mass-ejection-driven orbital evolution}

   \author{\begin{CJK}{UTF8}{gbsn}
Xiao-Tian Xu (徐啸天)\inst{\ref{TDLI}}
\end{CJK}
\and
Philipp Podsiadlowski\inst{\ref{LISA},\ref{Oxford}} 
\and
Norbert Langer\inst{\ref{AIfA}}
\and
Xue-Feng Wang\inst{\ref{PMO}}
\and
Xiang-Dong Li\inst{\ref{NJU1},\ref{NJU2}}
\and
Alexander Heger\inst{\ref{Monash}}
\and
Jonathan Mackey\inst{\ref{DIAS}}
\and
G\"otz Gr\"afener\inst{\ref{AIfA}}
\and
Harim Jin\inst{\ref{AIfA},\ref{MPIA}}
}

\institute{
Tsung-Dao Lee Institute, Shanghai Jiao-Tong University, 1 Lisuo Road, Shanghai, 201210, People's Republic of China \label{TDLI}\\
\email{xxu-tdli@sjtu.edu.cn}
\and
London Institute for Stellar Astrophysics, Vauxhall, London \label{LISA}
\and
University of Oxford, St Edmund Hall, Oxford, OX1 4AR, United Kingdom \label{Oxford}
\and
Argelander-Institut f\"ur Astronomie, Universit\"at Bonn, Auf dem H\"ugel 71, 53121 Bonn, Germany \label{AIfA}
\and
Purple Mountain Observatory, Chinese Academy of Sciences, Nanjing 210023, People's Republic of China\label{PMO}
\and
School of Astronomy and Space Science, Nanjing University, Nanjing, 210023, People's Republic of China\label{NJU1}
\and
Key Laboratory of Modern Astronomy and Astrophysics, Nanjing University, Ministry of Education, Nanjing, 210023, People's Republic of China\label{NJU2}
\and
School of Physics and Astronomy, Monash University, Vic 3800, Australia\label{Monash}
\and
Dublin Institute for Advanced Studies, Astronomy \& Astrophysics Section, DIAS Dunsink Observatory, Dublin, D15 XR2R, Ireland\label{DIAS}
\and
Max-Planck-Institut f\"ur Astrophysik, Karl-Schwarzschild-Strasse 1, 85748 Garching, Germany\label{MPIA}
}

\titlerunning{Mass-ejection-driven orbital evolution}

\date{Submitted date / Received date / Accepted date }

\abstract
{
Long-period Wolf-Rayet (WR) star binaries produced by mass transfer are predicted to be abundant, but are observationally rare. This yields constraints on the evolution of initially wide O star
binaries, including those potentially leading to the formation of gravitational-wave sources through the Common Envelope Channel.
}
{
We investigate this issue in the light of a new type of orbital evolution for initially wide O star binaries, which is
driven by mass ejection at periastron passage during the Luminous Blue Variable (LBV) phase.
}
{
The assumption that the mass ejection occurs instantly at periastron passage allows us to analytically describe the orbital evolution. This approach is motivated by our understanding of an Eddington-limit driven LBV phase. We perform population synthesis calculations for
the WR stars in the Small Magellanic Cloud (SMC), and compare them to
the observed SMC WR star population.
}
{
Different from mass transfer, our mass ejection scenario leads to increased orbital periods and eccentricities.
The Galactic system WR\,140 (orbital period 2895 d, eccentricity 0.9) could be a typical result of this evolution scenario. 
Our models predict measurable binary space velocities, and allow for the disruption of the binary.
Our SMC population synthesis model predicts statistically 5.3 close, 3.7 long-period, and further 2 runaway single WR stars.
With largely increased orbital periods and eccentricities,
such WR+O star binaries may not be ruled out by past radial-velocity searches. Applying our scenario to the Gaia BH1 and BH2 systems, we find that it provides viable progenitor evolution models.
}
{
The mass-ejection-driven orbital evolution could explain why so few wide WR binaries are observed, 
and why some of the apparently single WR stars have high space velocities.
We discuss implications for gravitational-wave sources.
}


\maketitle

\nolinenumbers

\section{Introduction\label{intro}}

\begin{figure*}[t!]
    \centering
    \includegraphics[trim=0mm 32mm 0mm 22mm, clip,width=\linewidth]{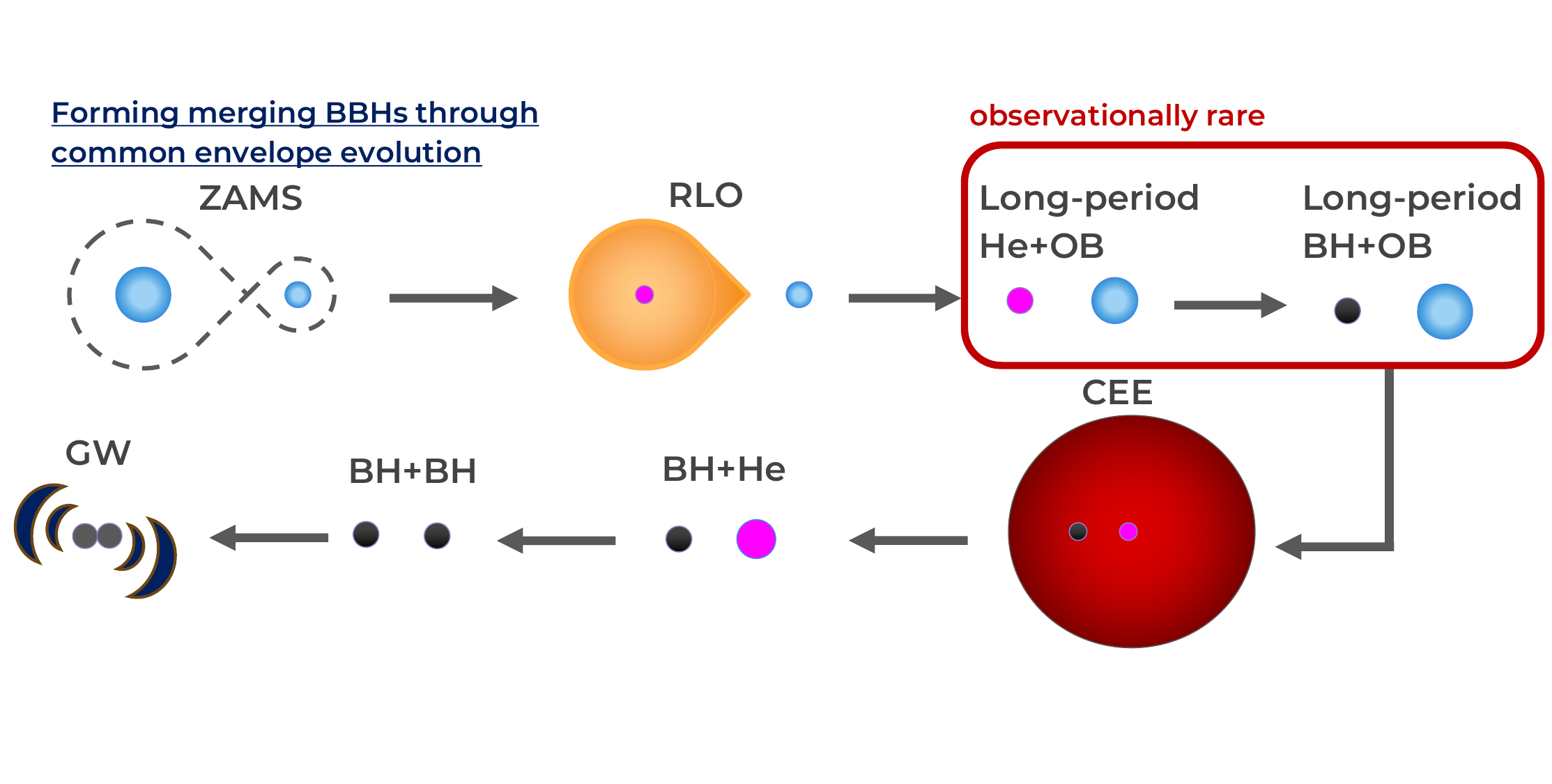}
    \caption{Schematic evolutionary sequence for forming merging black holes through common envelope evolution from isolated massive binaries. The meaning of the abbreviations are, 1) "ZAMS" for zero-age main sequence, 2) "RLO" for Roche-lobe overflow, 3) "He" for helium star, 4) "OB" for OB-type main-sequence star, 5) "CEE" for common envelope evolution, 6) "GW" for gravitational wave. Cumulating evidence shows that the progenitors (indicated by the red box) of the CEE phase are rarer than expected in recent surveys. }
    \label{fig:schematic}
\end{figure*}

Observations show that most massive stars have a nearby companion star \citep{Sana2012,Sana2013A&A...550A.107S,Moe2017ApJS..230...15M,Sana2025}. Hence, the evolution of isolated massive binaries may provide a natural way to produce merging binary black holes \citep[BBHs; e.g.,][]{Belczynski2016,Marchant2021,Mandel2022,Marchant2024ARA&A..62...21M,Klencki2025arXiv250508860K,Xu2025arXiv251220054X,Briel2026arXiv260203629B}. Several evolutionary channels have been identified, including the channel involving common envelope evolution \citep[CEE;][]{Ivanova2013,Lau2025A&A...699A.274L,Chen2025arXiv251014173C}. Whereas large uncertainties in stellar physics \citep{Langer2012} and mass transfer physics \citep{Marchant2024ARA&A..62...21M} impact the predicted merger rates of BBHs, the CEE channel may play a dominating role in forming merging BBHs due to the strong orbital shrinkage that naturally occurs in such systems \citep{Belczynski2016,Mapelli2020,Mandel2022}.  

Figure \ref{fig:schematic} shows a schematic plot of the CEE channel for merging BBHs. The expansion of the primary star triggers the first mass transfer phase. We refer to \citet{Eldridge2017PASA...34...58E}, \citet{Langer2020}, \citet{Wang2020}, \citet{Sen2022}, \citet{Fragos2023ApJS..264...45F}, \citet{Xu2025A&A...704A.218X}, \citet{Schurmann2025A&A...704A.219S}, and \citet{Jin2025} for detailed descriptions on the physics of the first mass transfer phase. The stripping of the hydrogen-rich envelope of the primary star exposes the massive helium core (He star), which can appear as a Wolf-Rayet (WR) star if massive enough\footnote{In this work, we only consider classical WR stars, which are core-helium-burning stars formed through binary-stripping or self-stripping \citep{Shenar2020}, while some WR stars are likely burning hydrogen \citep{Grafener2008A&A...482..945G,Shenar2024arXiv241004436S}.}. Then, WR stars at low metallicities usually end their lives as black holes \citep[BH;][]{Aguilera-Dena2023A&A...671A.134A}.
Since strong orbital shrinkage occurs in the following CEE phase, forming a merging BBH in the end requires the WR- and BH-main-sequence star binaries to be wide enough to avoid merging during the CEE phase
\citep{Xu2024thesis,Xu2025A&A...704A.218X}.

Whereas the CEE channel can reproduce various properties of the observed merging BBHs \citep{Belczynski2016,Kruckow2018,Bavera2021A&A...647A.153B,Broekgaarden2022ApJ...938...45B,Olejak2024A&A...689A.305O,Willcox2025arXiv251007573W}, there is an increasing number of studies suggesting that long-period WR/BH-MS binaries --- the progenitor systems of the CEE phase --- appear to be rare, at the metallicities of the Magellanic Clouds and the Milky Way \citep{Pauli2022A&A...667A..58P,Schootemeijer2024,Deshmukh2024A&A...692A.109D,Xu2025A&A...704A.218X}. 

For example, the observed sample of WR stars in the Small Magellanic Cloud (SMC) is thought to be complete, of which 7 appear single, 4 are WR+O binaries, and one is a WR+WR binary \citep{Foellmi2003,Shenar2016,Shenar2018,Schootemeijer2018}. Detailed binary evolution models well reproduce the observed four WR+O binaries\citep{Wang2024ApJ...975L..20W,Xu2025A&A...704A.218X}, which have orbital periods below 20\,d. 
At the same time, \citet{Xu2025A&A...704A.218X} predict a similar number of long-period (above 20\,d) WR+OB binaries. 
The majority of those would have been detected by the monitoring campaign of \citet{Schootemeijer2024}, which excludes companions to the apparently single WR stars more massive than $5\mso$ with periods below 1 year.
The Large Magellanic Cloud shows a similar lack of long-period massive evolved binaries \citep{Langer2020,Pauli2022A&A...667A..58P}. A few Galactic Wolf-Rayet star binaries are observed to have long orbital periods, but their number is below the expectation of population synthesis calculations \citep{Deshmukh2024A&A...692A.109D}.

One way to alleviate this problem would be to assume that most wide O\,star binaries merge during
their mass transfer. This could happen if the mass transfer efficiency would be higher than in the
models quoted above. When a mass gainer accretes a high fraction of the transferred mass, its envelope inflates \citep{Lau2024ApJ...966L...7L,Schurmann2024A&A...691A.174S,Henneco2024A&A...682A.169H,Wang2026}, which can trigger mass overflow through the outer Lagrangian point, causing the mass transfer to turn unstable \citep{Schurmann2024A&A...691A.174S,Henneco2024A&A...682A.169H,Scherbak2025ApJ...990..172S}. Population synthesis studies on Be X-ray binaries do suggest a mass transfer efficiency of about 50\% \citep{Shao2014,Vinciguerra2020,Schurmann2025A&A...704A.219S,Xing2026}, while a low  mass transfer efficiency seems to be more suitable for more massive systems \citep{Petrovic2005,Shao2016,Xu2025A&A...704A.218X,Schurmann2025A&A...704A.219S,Zapartas2025arXiv250812677Z}. \cite{Schurmann2025A&A...704A.219S} find that an increased mass transfer efficiency indeed predicts fewer long-period WR star binaries, but still expecting a considerable fraction of long-period systems. 

In this work, we investigate an alternative scenario for the evolution of wide massive O star binaries.
The Milky Way contains a small group of WR binaries with O8–O9 companions and one O5 companion (WR~140), featuring long orbital periods ($P_{\rm orb}$$\sim$10 yr) and detected through dust emission \citep{Rosslowe2015MNRAS.447.2322R,Williams2019MNRAS.488.1282W,Thomas2021}. Three of them have eccentricity ($e$) measurements, WR~140 \citep[$e$: 0.8993, $P_{\rm orb}$: 2895\,d;][]{Thomas2021},  WR~19 \citep[$e$: 0.8, $P_{\rm orb}$: 10.1\,yr;][]{Williams2009MNRAS.395.2221W}, and WR~137 \citep[$e$: 0.18, $P_{\rm orb}$: 13.05\,yr;][]{Lefevre2005MNRAS.360..141L}.
If such WR~140-like systems also exist in the Magellanic Clouds, they might remain undetected binaries based on past radial velocity variability searches \citep{Foellmi2003,Hainich2014A&A...565A..27H,Schootemeijer2024}. If they avoided the WC phase,
also episodic dust production would not occur.

At the same time, the observed properties of WR~140 can hardly be explained by current binary evolution models.  
The orbital separation of WR~140 at periastron is about $293\rso$, suggesting that past strong binary interaction was inevitable. When the progenitor of the WR star approaches Roche-lobe filling, current binary evolution models expect the orbit to be circularised efficiently \citep[e.g.,][]{Verbunt1995A&A...296..709V,Hurley2002,Belczynski2008,Kruckow2018,Fragos2023ApJS..264...45F}, and the following mass transfer phase does not induce eccentricities either \citep{Tauris2006,Tauris2023pbse.book.....T,Marchant2024ARA&A..62...21M}. It is also proposed that mass transfer may occur in eccentric binaries, which can enhance the eccentricity depending on the mass ratios and mass transfer efficiencies \citep{Sepinsky2007,Rocha_2025,Parkosidis2025arXiv250905243P,Parkosidis2025arXiv251107190P}. While this scenario can reproduce the observed eccentricities of evolved low-mass binaries \citep{Parkosidis2025arXiv250905243P}, it cannot explain the proper motion velocity of WR~140 \citep[$\sim34\kms$;][]{Dzib2009RMxAA..45....3D}.

As the primary stars in wide O\,star binaries can expand considerably, they may reach their Eddington limit and turn into LBVs before mass transfer is initiated.
We propose that the orbital evolution of such binaries is then driven by the mass ejection during the luminous blue variable (LBV) phase of the primary stars instead of the mass transfer through Roche-lobe overflow. While this has been suggested before \citep[e.g.,][]{Vanbeveren1991A&A...252..159V,Bavera2023NatAs...7.1090B,Willcox2025arXiv251007573W,Olejak2025arXiv251110728O}, it was applied to circular binaries. We will show that, if this mass ejection preferentially occurs at periastron, it efficiently increases orbital periods, enhances eccentricities, produces measurable space velocities, and even causes binary disruption if the pre-mass-ejection binary has a large eccentricity. 

This paper is organised as follows. We derive the equations for the mass-ejection-driven orbital evolution and introduce our method for population synthesis in Sect.\,\ref{sec:model}. In Sect.\,\ref{sec:results}, we present example models for the mass-ejection-driven orbital evolution, the application to WR~140, and population synthesis predictions for the WR stars in the SMC. We compare with observations and discuss the implications and uncertainties of our model in Sect.\,\ref{sec:discussion}. We summarise our work in Sect.\,\ref{sec:conclusion}.

\section{Model and method\label{sec:model}}

In this section, we derive the equations for the mass-ejection-driven orbital evolution (Sect.\,\ref{sec:me-drive-model}). Then, we introduce our method for population synthesis calculations (Sect.\,\ref{sec:popsync-method}).

\subsection{Mass ejection driven orbital evolution\label{sec:me-drive-model}}

We assume that the mass-ejection-driven orbital evolution takes place when the primary star enters its LBV phase. 
As at this stage the binary orbit is likely to still be eccentric, we expect the primary will approach Roche lobe filling at periastron passage. Here, we assume that the loosely bound inflated envelope of the primary star can be kicked off instantly at periastron passage, with tidal effects on the orbital period and separation being negligible. We discuss in detail the trigger mechanism of this periastron mass ejection in Sect.\,\ref{sec:dis-trigger}. A similar process has been proposed for asymptotic giant branch stars, whose mass loss rate can be enhanced by tidal interaction in binaries, producing eccentricities \citep{Soker2000A&A...357..557S,Marinovic2008A&A...480..797B,Vos2015A&A...579A..49V}.
While our assumption is certainly an oversimplification of the situation, it allows an analytic prescription of the further evolution which might still capture the main features in such events. For primary stars which are not massive enough to reach the LBV phase, we assume that its dense envelope allows for tidal circularisation and ordinary Roche-lobe overflow evolution.

We first investigated the dynamical effect of one sudden mass ejection at periastron, which has been discussed previously \citep{Blaauw1961,Hills1983ApJ...267..322H,Tauris1998,vdH2000A&A...364..563V,Hurley2002,Pfahl2002ApJ...573..283P}. 
In this work, we extended the analysis by \citet{Hills1983ApJ...267..322H}, and we calculated the orientation of the post-mass-ejection orbit with the coordinate-independent description provided in \citet{Pfahl2002ApJ...573..283P}. Then, we establish the equations for long-term orbital evolution. For completeness, we also present the orbital-evolution equations for mass ejections occurring at apastron in App.\,\ref{app:apastron_case}.

\subsubsection{Pre-mass-ejection orbit}

\begin{figure}
    \centering
    \includegraphics[width=\linewidth]{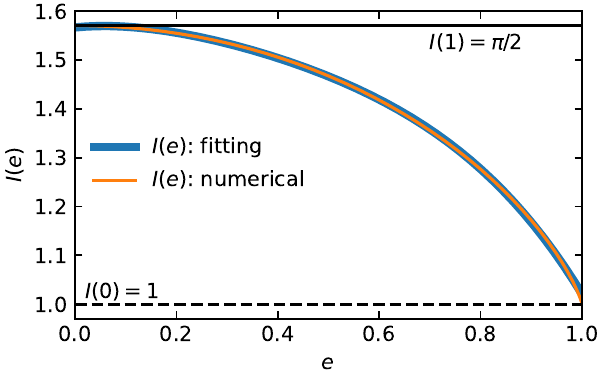}
    \caption{Comparison between our fitting formula (blue curve; Eq.\,\ref{eq:Ie-fit}) and the numerical solution of Eq.\,\eqref{eq:Ie} (orange curve). The black solid and dashed lines represent $I(1)=\pi/2$ and $I(0)=1$. }
    \label{fitting-I(e)}
\end{figure}

The two-body problem (i.e., the Kepler problem) well describes the properties of the pre-mass-ejection orbit.
Considering a binary system with stellar components having masses of $M_{\rm s1}$ ("s1" for star\,1, the primary star) and $ M_{\rm s2}$ ("s2" for star\,2, the secondary star), the velocity of the centre of mass, $\upsilon_{\rm CoM,0}$, is defined as 
\begin{equation}
    M_{\rm s1}  \vv{\upsilon}_{\rm s1} + M_{\rm s2} \vv{\upsilon}_{\rm s2}=M_{\rm b,0} \vv{\upsilon}_{\rm CoM,0},
\end{equation}
where $M_{\rm b,0}$ is the total mass of the binary (i.e., $M_{\rm s1}+M_{\rm s2}$), where the subscript "0" indicates the value before the mass ejection event,  ${\upsilon}_{\rm s1}$ and ${\upsilon}_{\rm s1}$ are the velocities of star\,1 and star\,2, and the relative velocity, $\upsilon_0$, is defined as
\begin{equation}
    \vv{\upsilon}_{\rm s1}  -\vv{\upsilon}_{\rm s2} = \vv{\upsilon}_{0},
\end{equation}
and the space velocities of the stellar components are given by 
\begin{equation}
    \vv{\upsilon}_{\rm s1}= \vv{\upsilon}_{\rm CoM,0} + \frac{M_{\rm s2}}{M_{\rm b,0}}\vv{\upsilon_0}, 
    \label{vs1}
\end{equation}
and 
\begin{equation}
    \vv{\upsilon}_{\rm s2} = \vv{\upsilon}_{\rm CoM,0} - \frac{M_{\rm s1}}{M_{\rm b,0}}\vv{\upsilon_0}.
    \label{vs2}
\end{equation}
The orbital energy, $E_0$, is given by 
\begin{equation}
    E_0=-G\frac{M_{\rm s1} M_{\rm s2}}{r} + \frac{1}{2} \mu_0 \upsilon_0^2 = - G\frac{M_{\rm s1} M_{\rm s2}}{2a_0},
    \label{E0}
\end{equation}
or equivalently
\begin{equation}
    \frac{E_0}{\mu_0}=-G\frac{M_{\rm b,0}}{r} + \frac{1}{2}\upsilon_0^2 = - G\frac{M_{\rm b,0}}{2a_0},
\end{equation}
where $G$ is the gravitational constant, $r$ is the orbital separation, $\mu_0$ is the reduced mass, which is $\mu_0 = M_{\rm s1} M_{\rm s2}/M_{\rm b,0}$, and $a_0$ is the semi-major axis.  The orbital angular momentum, $J_0$, is given by 
\begin{equation}
|\vv{J}_0|=\mu_0|\vv{r}\times \vv{\upsilon}_0|=\mu_0 \sqrt{GM_{\rm b,0}\,a_0(1-e_0^2)},
\label{J0}
\end{equation}
where $e_0$ is the eccentricity. The orbital separation can also be expressed as a function of true anomaly, $\theta$, which is
\begin{equation}
    r=\frac{a_0(1-e_0^2)}{1+e_0\cos\theta}.
    \label{eq:r-anomaly}
\end{equation}
The separation at periastron is evaluated by taking $\theta = 0$, 
\begin{equation}
    r_{\rm per,0} = a_0 (1-e_0).
    \label{per1}
\end{equation}
Combining Eqs.\,\eqref{per1} with \eqref{E0}, we obtained the relative velocity at periastron,
\begin{equation}
    \upsilon_{\rm per,0} = \sqrt{\frac{GM_{\rm b,0}}{a_0}\frac{1+e_0}{1-e_0}} = \left(\frac{GM_{\rm b,0}}{P_{\rm orb,0}}2\pi\right)^{1/3}\sqrt{\frac{1+e_0}{1-e_0}},
    \label{eq:vper0}
\end{equation}
where  $a_0$ was replaced by the orbital period, $P_{\rm orb,0}$, through the Kepler's third law, 
\begin{equation}
    \left(\frac{2\pi}{P_{\rm orb,0}}\right)^2 a_0^3=GM_{\rm b,0}.
\end{equation}
In addition, we introduced an additional parameter to measure the detectability of a given binary system through radial velocity measurements, which is the orbital-period-averaged relative velocity,
\begin{equation}
\begin{split}
    \langle \upsilon_{\rm orb} \rangle &= \int_{0}^{P_{\rm orb}} \frac{\upsilon_{\rm 0}}{P_{\rm orb,0}}\dd t = \int_{0}^{2\pi} \frac{\upsilon_{\rm 0}}{P_{\rm orb,0}}\frac{\dd t}{\dd \theta}\dd \theta\\
    &= \int_{0}^{2\pi} \frac{\upsilon_{\rm 0}}{P_{\rm orb,0}\omega_{\rm orb}}\dd \theta,
    \label{eq:vorb0bar}
    \end{split}
\end{equation}
where $t$ is the evolutionary time, $\theta$ is the true anomaly in Eq.\,\eqref{eq:r-anomaly}, and $\omega_{\rm orb}$ is the Keplerian angular velocity, which is 
\begin{equation}
      \omega_{\rm orb} = \frac{\sqrt{ GM_{\rm b,0}\,a_0\,(1-e_0^2)}}{r^2}.
      \label{eq:w-orb}
\end{equation}
Combining Eqs.\,\eqref{E0}, \eqref{eq:r-anomaly}, and \eqref{eq:w-orb} with Eq.\,\eqref{eq:vorb0bar}, we obtained 
\begin{equation}
\begin{split}
    \langle \upsilon_{\rm orb} \rangle &=2\frac{a_0(1-e_0^2)}{P_{\rm orb,0}}
    \int_{0}^{\pi} \sqrt{\frac{1+2e_0\cos\theta+e_0^2}{(1+e_0\cos\theta)^4}} \dd \theta\\
    &=4\frac{a_0}{P_{\rm orb,0}}
    \int_{0}^{\pi/2} \sqrt{1-e_0^2\sin^2\phi}\, \dd \phi \\
    &=4\frac{a_0}{P_{\rm orb,0}} I(e_0)
    = 4\left[\frac{GM_{\rm b,0}}{4\pi^2P_{\rm orb,0}}\right]^{1/3}I(e_0),
\end{split}
\end{equation}
where the elliptic integral,
\begin{equation}
    I(e_0) = \int_{0}^{\pi/2} \sqrt{1-e_0^2\sin^2\phi}\, \dd \phi,
    \label{eq:Ie}
\end{equation}
cannot be solved analytically \citep{Murray1999ssd..book.....M}. Hence, we applied an empirical formula to fit the numerical solution of Eq.\,\eqref{eq:Ie} \citep{Abramowitz1972hmfw.book.....A}, which is
\begin{equation}
    I(e)\simeq C_1 + C_2\,e + C_3\,e^2 + C_4\,e^3+C_5\,e^4,
    \label{eq:Ie-fit}
\end{equation}
where
\begin{equation}
\begin{cases}
C_1= 1.56690314\\ 
C_2=  0.10180252\\
C_3= -0.97975648\\
C_4=1.16831992\\
C_5= -0.83659936
\end{cases}.
\end{equation}
The maximal relative error (i.e., $|\text{numerical}-\text{fitting}|/\text{numerical}$) of this fitting is $2\%$. Figure \ref{fitting-I(e)} compares our fitting formula and the numerical solution. Then, the averaged orbital velocity in the centre-of-mass frame of the mass-losing star is given by
\begin{equation}
    \langle\upsilon_{\rm s1}\rangle = \frac{M_{\rm s2}}{M_{\rm b,0}}\langle \upsilon_{\rm orb} \rangle.
\end{equation}
In addition, the orbital velocities in the centre-of-mass frame of the mass-losing star at periastron and apastron are
\begin{equation}
    \upsilon_\text{s1,per}=\frac{M_\text{s2}}{M_{\rm b,0}}\upsilon_{\rm per,0}
    \label{eq:v-s1-per}
\end{equation}
and
\begin{equation}
    \upsilon_\text{s1,ap}=\frac{M_\text{s2}}{M_{\rm b,0}}\upsilon_{\rm ap,0},
    \label{eq:v-s1-ap}
\end{equation}
where $\upsilon_{\rm ap,0}$ is the relative velocity at apastron (Eq.\,\ref{eq:v-ap0}).

\subsubsection{Post-mass-ejection orbit\label{sec:post-ME-orbit}}

For analytical simplicity, we adopted the following assumptions on the sudden mass ejection at periastron, consistent with previous works \citep{Hills1983ApJ...267..322H,Tauris1998,Pfahl2002ApJ...573..283P}, and we discuss the caveats in Sect.\,\ref{sec:caveats}:
\begin{enumerate}
    \item the mass ejection is isotropic with respect to the primary star and does not induce additional momentum kicks on the mass-losing star (see \citealt{Hills1983ApJ...267..322H}, \citealt{Tauris1998}, and \citealt{Pfahl2002ApJ...573..283P} for the effect of significant momentum kicks in supernova explosions);
    \item the mass ejection is a delta function in time (cf. Sect.\,\ref{sec:dis-trigger});
    \item the effect of the tidal torque is ignorable;
    \item the velocity of the centre of mass of the pre-mass-ejection system is zero (i.e., $\upsilon_{\rm CoM,0}=0$);
    \item the binary system does not undergo apsidal advance or precession induced for external reasons (cf. Sect.\,\ref{sec:dis-prece} and App.\,\ref{app:precession}).
\end{enumerate}

We assumed that the ejection event takes away a mass of $\Delta M$, while the mass of the secondary star remains unchanged. According to our sudden ejection assumption, the relative velocity, space velocities\footnote{The orbital velocities in the centre-of-mass frame do change because of the changes in masses (Eqs. \ref{eq:v-s1-per} and \ref{eq:v-s1-ap}). It can be shown that 
\begin{equation*}
    \vv{\upsilon}_{\rm s1} =\frac{M_{\rm s2}}{M_{\rm b,0}}\vv{\upsilon}_{\rm per,0} = \frac{M_{\rm s2}}{M_{\rm b,0}-\Delta M}\vv{\upsilon}_{\rm per,0} + \vv{\upsilon}_{\rm CoM,1},
\end{equation*}
which is consistent with our assumption.
}, and orbital separation remain unchanged right after the ejection. Hence the post-mass-ejection orbital energy is given by 
\begin{equation}
    -G\frac{M_{\rm b,0}-\Delta M}{r_{\rm per,0}} + \frac{1}{2} \upsilon_{\rm per,0}^2 = - G\frac{M_{\rm b,0}-\Delta M}{2a_1},
    \label{E1}
\end{equation}
where $a_1$ is the semi-major axis of the post-mass-ejection orbit, and the subscript "1" indicates the value after mass ejection. The post-ejection orbital angular momentum is
\begin{equation}
\begin{split}  
    J_1&=\mu_1 |\vv{r}_{\rm per,0} \times \vv{\upsilon}_{\rm per,0} | 
    &=\mu_1 \sqrt{G(M_{\rm b,0}-\Delta M)a_1(1-e_1^2)}, 
    \label{J1}
\end{split}
\end{equation}
where $\mu_1=(M_{\rm s1} - \Delta M)M_{\rm s2}/(M_{\rm b,0}-\Delta M)$, and $e_1$ is the post-mass-ejection eccentricity. The post-mass-ejection orbital period, $P_{\rm orb,1}$, is given by Kepler's third law,
\begin{equation}
    P_{\rm orb,1}^2 = \frac{(2\pi)^2 a_1^3}{G(M_{\rm b,0}-\Delta M)}.
    \label{P1}
\end{equation}
Combining the above equations, we obtained the properties of the post-mass-ejection orbit, 
\begin{equation}
    a_1=a_0 (1-e_0) \frac{M_{\rm b,0}-\Delta M}{M_{\rm b,0}(1-e_0)-2 \Delta M},
    \label{aper1}
\end{equation}
\begin{equation}
    e_1=\frac{\Delta M+e_0 M_{\rm b,0}}{M_{\rm b,0} - \Delta M},
    \label{eper1}
\end{equation}
\begin{equation}
    P_{\rm orb,1}^2 =P_{\rm orb,0}^2\frac{(1-\Delta M/M_{\rm b,0})^2(1-e_0)^3}{[(1-e_0)-2\Delta M / M_{\rm b,0}]^3}.
    \label{Pper1}
\end{equation}

Due to linear momentum conservation, the change of mass causes a momentum kick on the centre of mass,
\begin{equation}
     (M_{\rm s1}-\Delta M)  \vv{\upsilon}_{\rm s1} + M_{\rm s2} \vv{\upsilon}_{\rm s2}=(M_{\rm b,0}-\Delta M) \vv{\upsilon}_{\rm CoM,1},
\end{equation}
which gives the post-mass-ejection velocity of the centre of mass,
\begin{equation}
    \vv{\upsilon}_{\rm CoM,1} = \vv{\upsilon}_{\rm CoM,0} -\frac{\Delta M M_{\rm s2}}{M_{\rm b,0} (M_{\rm b,0}-\Delta M)}\vv{\upsilon}_{\rm per,0},
    \label{eq:vcom1}
\end{equation}
which shows that the momentum kick has the opposite direction to $\vv{\upsilon}_\text{per,0}$. As we assumed that the initial velocity of the centre of mass is zero, the magnitude of $\vv{\upsilon}_{\rm CoM,1}$ is 
\begin{equation}
    \upsilon_{\rm CoM,1} =\frac{\Delta M M_{\rm s2}}{M_{\rm b,0} (M_{\rm b,0}-\Delta M)}\upsilon_{\rm per,0}.    \label{vcom1}
\end{equation}

These equations imply that a mass ejection at periastron makes the post-mass-ejection orbit wider and more eccentric. 
Setting $e_1=1$, we find that the amount of the ejected mass required for binary disruption is 
\begin{equation}
 \Delta M_{\rm dis} = \frac{1}{2}M_{\rm b,0}(1-e_0),   
 \label{dm_per_disr}
\end{equation}
which is consistent with \citet{Hills1983ApJ...267..322H}. In addition, we note that the post-mass-ejection orbit meets the following condition,
\begin{equation}
    a_1 (1-e_1) = r_{\rm per,0}= a_0 (1-e_0),
    \label{eq:r-per}
\end{equation}
which means that the separation at periastron remains unchanged. While the orbit of the binary is no longer closed due to the mass ejection, the orientation of the post-mass-ejection orbit stays the same as the pre-mass-ejection orbit. In the following, we demonstrate this by considering the Laplace-Runge-Lenz vector\footnote{The definition is given by Eq.\,(B1) in \citet{Pfahl2002ApJ...573..283P}.}, which is
\begin{equation}
\begin{split}
    \vv{e}_0 & = \frac{\vv{\upsilon}_{\rm per,0}\times(\vv{r}_{\rm per,0}\times\vv{\upsilon}_{\rm per,0})}{GM_{\rm b,0}} - \frac{\vv{r}_{\rm per,0}}{r_{\rm per,0}}\\
    & = \frac{\upsilon_{\rm per,0}^2\vv{r}_{\rm per,0}-(\vv{r}_{\rm per,0}\,\cdot\,\vv{\upsilon}_{\rm per,0})\vv{\upsilon}_{\rm per,0}}{GM_{\rm b,0}} - \frac{\vv{r}_{\rm per,0}}{r_{\rm per,0}},
    \label{eq:LRL-vector}
\end{split}    
\end{equation}
where $\vv{r}_{\rm per,0}$ is the vector pointing from the secondary star to the mass-losing star. At periastron, $\vv{r}_{\rm per,0}$ is perpendicular to $\vv{\upsilon}_{\rm per,0}$, which gives
\begin{equation}
    \vv{r}_{\rm per,0}\,\cdot\,\vv{\upsilon}_{\rm per,0}=0.
\end{equation}
Hence, the Laplace-Runge-Lenz vector before the mass ejection is 
\begin{equation}
    \vv{e}_0 = \frac{\upsilon_{\rm 0,per}^2\vv{r}_{\rm per,0}}{GM_{\rm b,0}} - \frac{\vv{r}_{\rm per,0}}{r_{\rm per,0}}=\left[\frac{\upsilon_{\rm per,0}^2}{GM_{\rm b,0}} - \frac{1}{r_{\rm per,0}}\right]\vv{r}_{\rm per,0},
\end{equation}
which has the same orientation as $\vv{r}_{\rm per,0}$, as expected. After the mass ejection, the Laplace-Runge-Lenz vector becomes
\begin{equation}
    \vv{e}_1 = \left[\frac{\upsilon_{\rm per,0}^2}{G(M_{\rm b,0}-\Delta M)} - \frac{1}{r_{\rm per,0}}\right]\vv{r}_{\rm per,0}.
\end{equation}
It is clear that $M_{\rm b,0}-\Delta M < M_{\rm b,0}$, which leads to the following strict inequality,
\begin{equation}
\begin{split}
    \frac{\upsilon_{\rm per,0}^2}{G(M_{\rm b,0}-\Delta M)} - \frac{1}{r_{\rm per,0}} &> \frac{\upsilon_{\rm per,0}^2}{GM_{\rm b,0}} - \frac{1}{r_{\rm per,0}}\\
    &= \frac{e_0}{a_0(1-e_0)} > 0.
    \end{split}
\end{equation}
Therefore, we obtained 
\begin{equation}
\begin{cases}
    \vv{e}_0\times\vv{e}_1 \propto \vv{r}_{\rm per,0}\times\vv{r}_{\rm per,0}=0\\
    \vv{e}_0\,\cdot\,\vv{e}_1 = | \vv{e}_0\,\cdot\,\vv{e}_1|
\end{cases},
    \label{eq:direction-per}
\end{equation}
which means that the orientation of the post-mass-ejection orbit has the same orientation as the pre-mass-ejection orbit\footnote{If the mass ejection occurs at apastron, the pre-mass-ejection apastron can become the periastron of the post-mass-ejection orbit (i.e., $\vv{e}_0\,\cdot\,\vv{e}_1 = -| \vv{e}_0\,\cdot\,\vv{e}_1|$; see App.\,\ref{app:apastron_case}), depending on the ejected mass.}.

\subsubsection{Long-term orbital evolution\label{sec:long-term-evo}}

Based on our result on one mass ejection at periastron, in this section we derive the orbital-evolution equations by considering that the LBV primary star experiences a series of mass ejection events associated with periastron passages, and the amount of ejected mass during each periastron passage was fixed to be $\Delta M$. Unlike the approach in \citet{Sepinsky2007} (i.e., averaging equations over one orbital period), we described the long-term evolution by relating the orbital elements after $n$th and after $(n+1)$th periastron passage (see App.\,\ref{app:math-induction} for mathematical details).

Letting $(a_1,\,e_1,\,P_{\rm orb,1},\,\upsilon_{\rm CoM,1})$ be the orbital parameters after the first periastron passage, and the binary mass becomes $(M_\text{b,0}-\Delta M)$. Since the separation and orientation of periastron remains unchanged (Eqs.\,\ref{eq:r-per} and \ref{eq:direction-per}), we can obtain the parameters after the second mass ejection, $(a_2,\,e_2,\,P_{\rm orb,2},\,\upsilon_{\rm CoM,2})$, by replacing $M_{\rm b,0}$ in Eqs.\,\eqref{aper1}--\eqref{Pper1} and \eqref{vcom1} with $(M_{\rm b,0}-\Delta M)$, 
\begin{equation}
\begin{split}
    a_2&=a_1 (1-e_1) \frac{(M_{\rm b,0}-\Delta M)-\Delta M}{(M_{\rm b,0}-\Delta M)(1-e_1)-2\Delta M}\\
    &=a_0 (1-e_0) \frac{M_{\rm b,0}-2\Delta M}{M_{\rm b,0}(1-e_0)-2(2\Delta M)},
    \label{eq:a-2ndME}
\end{split}
\end{equation}
\begin{equation}
\begin{split}
    e_2&=\frac{\Delta M+e_1 (M_{\rm b,0}-\Delta M)}{(M_{\rm b,0}-\Delta M) - \Delta M}\\
    &=\frac{(2\Delta M)+e_0 M_{\rm b,0}}{M_{\rm b,0} - (2\Delta M)},    
\end{split}
\end{equation}
\begin{equation}
    \begin{split}
    P_{\rm orb,2}^2 &=P_{\rm orb,1}^2\frac{[1-\Delta M/(M_{\rm b,0}-\Delta M)]^2(1-e_1)^3}{[(1-e_1)-2\Delta M / (M_{\rm b,0}-\Delta M)]^3}\\
    &=P_{\rm orb,0}^2\frac{(1-2\Delta M/M_{\rm b,0})^2(1-e_0)^3}{[(1-e_0)-2(2\Delta M) / M_{\rm b,0}]^3},
    \end{split}
\end{equation}
\begin{equation}
    \begin{split}
    \vv{\upsilon}_{\rm CoM,2} &=\vv{\upsilon}_{\rm CoM,1} - \frac{\Delta M M_{\rm s2}}{(M_{\rm b,0}-\Delta M) [(M_{\rm b,0}-\Delta M)-\Delta M]}\vv{\upsilon}_{\rm per,0}.
    \end{split}
\end{equation}
According to Eq.\,\eqref{eq:vcom1}, the direction of $\vv{\upsilon}_{\rm CoM,1}$ is opposite to that of $\vv{\upsilon}_{\rm per,0}$, as $\upsilon_{\rm CoM,0}=0$ according to our assumption. Hence, the magnitude of $\vv{\upsilon}_{\rm CoM,2}$ is 
\begin{equation}
    \begin{split}
    \upsilon_{\rm CoM,2} &=\upsilon_{\rm CoM,1} + \frac{\Delta M M_{\rm s2}}{(M_{\rm b,0}-\Delta M) [(M_{\rm b,0}-\Delta M)-\Delta M]}\upsilon_{\rm per,0}\\
    &=\frac{(2\Delta M) M_{\rm s2}}{M_{\rm b,0} (M_{\rm b,0}-2\Delta M)}\upsilon_{\rm per,0}.\\
    \end{split}
    \label{eq:vcom-2ndME}
\end{equation}
The above equations are consistent with the expectation of Eqs.\,\eqref{aper1}--\eqref{Pper1} and \eqref{vcom1} if the ejected mass is taken to be $2\Delta M$ during one periastron passage, which means that the orbital parameters are determined by the initial orbital parameters, $(a_0,\,e_0,\,P_{\rm orb,0},\,\upsilon_{\rm CoM,0})$, and the total amount of ejected mass. If the binary stays bound, this is also true for the parameters after the $n$th periastron passage, $(a_n,\,e_n,\,P_{\rm orb,n},\,\upsilon_{\rm CoM,n})$,  which are
\begin{equation}
    a_n=a_0 (1-e_0) \frac{M_{\rm b,0}-n\Delta M}{M_{\rm b,0}(1-e_0)-2 n\Delta M},
    \label{eq:aper2}
\end{equation}
\begin{equation}
    e_n=\frac{n\Delta M+e_0 M_{\rm b,0}}{M_{\rm b,0} - n\Delta M},
    \label{eq:eper2}
\end{equation}
\begin{equation}
    P_{\text{orb},n}^2 =P_{\rm orb,0}^2\frac{(1-n\Delta M/M_{\rm b,0})^2(1-e_0)^3}{[(1-e_0)-2n\Delta M / M_{\rm b,0}]^3},
    \label{eq:Pper2}
\end{equation}
\begin{equation}
    \upsilon_{\text{CoM},n} =\frac{n\Delta M M_{\rm s2}}{M_{\rm b,0} (M_{\rm b,0}-n\Delta M)}\upsilon_{\rm per,0},
    \label{eq:vcomper1}
\end{equation}
where $n\Delta M$ represents the total ejected mass. In App\,\ref{app:math-induction}, we prove this using mathematical induction. Hence, if a binary enters the mass-ejection-driven orbital evolution, the orbital parameters at the end of the evolution are determined by the mass of the hydrogen-rich envelope of the LBV primary star, instead of the history of the mass ejections. For example, we envision that an LBV primary ejects $10^{-3}\mso$ during the first periastron passage and $2\times10^{-3}\mso$ during the second. The resulting orbital configuration is the same as the case where the LBV primary ejects $3\times10^{-3}\mso$ during a single periastron passage. 

Correspondingly, the orbital period, space velocity, average orbital velocity, and eccentricity of the resulting WR+O binary should be correlated through the ejected mass, which leads to the following relations,
\begin{equation}
    P_{\text{orb},n}^2=P_{\rm orb,0}^2\left(\frac{1+e_n}{1+e_0}\right)\left(\frac{1-e_0}{1-e_n}\right)^3,
    \label{eq:e-porb}
\end{equation}
\begin{equation}
    \upsilon_{\text{CoM},n} = \frac{M_{\rm s2}}{M_{\rm b,0}}\left(\frac{e_n-e_0}{1+e_0}\right)\upsilon_{\rm per,0},
    \label{eq:vcom-e}
\end{equation}
and
\begin{equation}
    \langle\upsilon_{\rm orb}\rangle = 4\left[\frac{GM_{\text{b},n}}{4\pi^2P_{\rm orb,0}}\right]^{1/3} I(e_n)\left(\frac{1+e_0}{1+e_n}\right)^{1/6}\left(\frac{1-e_n}{1-e_0}\right)^{1/2}.
    \label{eq:e-vorb}
\end{equation}

Binary disruption occurs when orbital energy is equal to or exceeds zero (i.e., $e_{\rm n}\geq1$). In the following, we derive the runaway velocities of the stellar components, which are defined as the space velocities when the separation of two components is infinite. According to the equations in \citet{Pfahl2002ApJ...573..283P}, the runaway velocities, $\vv{\upsilon}_{\rm s1,\infty}$ and $\vv{\upsilon}_{\rm s2,\infty}$, are evaluated by 
\begin{equation}
    \vv{\upsilon}_{\rm s1,\infty} = \frac{M_{\rm s2,dis}}{M_{\rm b,dis}}\vv{\upsilon}_{\rm \infty} +\vv{\upsilon}_{\rm CoM,dis},
    \label{eq:vrun1}
\end{equation}
and 
\begin{equation}
    \vv{\upsilon}_{\rm s2,\infty} = -\frac{M_{\rm s1,dis}}{M_{\rm b,dis}}\vv{\upsilon}_{\rm \infty} +\vv{\upsilon}_{\rm CoM,dis},
    \label{eq:vrun2}
\end{equation}
where "dis" indicates that the parameters are evaluated at disruption, and $\vv{\upsilon}_{\rm \infty}$ is the relative velocity at infinite separation. In the case of a supernova explosion, the asymmetric ejecta generates a strong momentum kick on the newborn compact object \citep{Janka2017}. Consequently, the remnant of the exploding star moves at a velocity close to the kick velocity, while the companion star has a velocity close to its orbital velocity at explosion \citep{Hills1983ApJ...267..322H,Tauris1998,Pfahl2002ApJ...573..283P}. However, binary disruption occurs in a different way in our mass-ejection-driven orbital evolution where the eccentricity increases gradually with each periastron passage. When disruption occurs, the eccentricity should be equal to or slightly above 1, and the orbital energy should be very close to zero. Then, in the centre-of-mass frame, we have the following relation for the kinetic energy, $E_{\rm kin}$, and potential energy, $E_{\rm grav}$, of the system at disruption, 
\begin{equation}
    E_{\rm kin} + E_{\rm grav} = 0,
\end{equation}
which suggests that, at infinite separation (i.e., $E_{\rm grav} = 0$), the relative velocity is zero, and, according to Eqs.\,\eqref{eq:vrun1} and \eqref{eq:vrun2}, the runaway velocities of both stars are equal to the velocity of the centre of mass at disruption, $\vv{\upsilon}_{\rm CoM,dis}$. By inserting Eq.\,\eqref{dm_per_disr} into Eq.\,\eqref{eq:vcomper1}, or setting $e_n=1$ in Eq.\,\eqref{eq:vcom-e}, we obtained
\begin{equation}
   \upsilon_{\rm CoM,dis} =\frac{M_{\rm s2}}{M_{\rm b,0}} \left(\frac{1-e_0}{1+e_0}\right)\upsilon_{\rm per,0}=\upsilon_{\rm s1,\infty}= \upsilon_{\rm s2,\infty}.
   \label{eq:vcomdis}
\end{equation}

\subsection{Population synthesis method\label{sec:popsync-method}}

Our population synthesis calculations were based on Eqs.\,\eqref{eq:aper2}-\eqref{eq:vcomper1}, and Eq.\,\eqref{eq:vcomdis}, where $n\Delta M$ was set equal to 
the hydrogen-rich envelope of the primary star, $M_{\rm b,0}$ is taken to be the binary mass and $P_{\rm orb,0}$ the orbital period on the zero-age main sequence (ZAMS), respectively. We ignored the effects of steady stellar winds. 
We adopted a dense grid of detailed binary evolution models computed with a metallicity tailored for the SMC and the SMC WR star population derived from it as inputs \citep[see][for the details of the binary evolution models]{Xu2025A&A...704A.218X}, where the WR stars were defined as core-helium-burning stars with luminosities above $10^{5.6}\lso$ based on observations \citep{Shenar2020}. We adopted the WR star mass, $M_{\rm WR}$, derived by \citet{Xu2025A&A...704A.218X}, and calculated the mass of the hydrogen-rich envelope by $n\Delta M = M_{\rm ZAMS} - M_{\rm WR}$, where $M_{\rm ZAMS}$ is the ZAMS mass of the WR star.

In order to determine the occurrence of the mass-ejection-driven orbital evolution, we introduced the following criterion. The mass-ejection-driven orbital evolution is assumed to occur if the initial primary mass exceeds a critical value, $M_{\rm crit}$, and simultaneously the initial orbital period is longer than a critical value, $P_{\rm crit}$. The physical motivation is that, for mass ejection to occur, the primary star needs to be massive enough to enter an LBV phase by reaching its Eddington limit, and the orbit needs to be wide enough to contain an inflated star.
The chosen parameters are approximately reflecting the Humphreys-Davidson limit (cf., Sect.\,\ref{sec:caveats}).  We addressed the uncertainties by varying the values for $M_{\rm crit}$ and $P_{\rm crit}$ (Sect.\,\ref{sec:param-study} and App.\,\ref{app:param-study})

For our Simulation1, we take $M_{\rm crit}=30\mso$ and $P_{\rm crit}=40\,{\rm d}$. The luminosity of a $30\mso$ star at the terminal-age main sequence is about $10^{5.4}\lso$ \citep{Xu2025A&A...704A.218X}, which is consistent with the observed minimal luminosity of LBVs \citep{Smith2004ApJ...615..475S,Vink2012ASSL..384..221V,Smith2017RSPTA.37560268S,Smith2019MNRAS.488.1760S}, and the adopted $40\,$d for $P_{\rm crit}$ is consistent with the inferred orbital period of the progenitor of WR~140 (cf., Sect.\,\ref{sec:wr140}).
For systems in which the above criteria are not met, we adopt the orbital periods for WR+O binaries obtained by \citet{Xu2025A&A...704A.218X}.

Consistent with the fiducial population synthesis model in \citet{Xu2025A&A...704A.218X}, we assumed that the initial primary mass, $\mi$, follows the initial mass function, 
\begin{equation}
    f_{\rm IMF} \propto \mi^{-\alpha},
\end{equation}
where $\alpha=2.3$ for massive stars \citep{Kroupa2001}, and we adopted the initial mass ratio, $f_{\qi}$, and orbital period, $f_{\logpi}$, functions in \citet{Sana2012}, which are 
\begin{equation}
    f_{\qi} \propto \qi^{-0.1},
\end{equation}
and 
\begin{equation}
    f_{\logpi} \propto (\logpi)^{-0.55}.
    \label{eq:IPF}
\end{equation}
In addition to \citet{Xu2025A&A...704A.218X}, we further took into account the initial eccentricity function, 
\begin{equation}
    f_e \propto e^{-\eta}
\end{equation}
We take $\eta=0.5$ for our Simulation1 according to the empirical eccentricity function in  \citet{Sana2012}. The normalisation factor of the $f_e$ function is $(-\eta+1)$. Hence the fraction of the binaries having eccentricities in $[e_1,\,e_2]$ is given by
\begin{equation}
    \int_{e_1}^{e_2}f_e\,\dd e=e_2^{-\eta+1} - e_1^{-\eta+1}.
    \label{eq:fe-fraction}
\end{equation}
The predicted number of each binary system was evaluated by a statistical weight, 
\begin{equation}
    N \propto \text{SFR}\,\times\,\text{lifetime}\,\times\, f_{\rm IMF}f_{\qi}f_{\logpi}f_e,
    \label{eq:weight}
\end{equation}
where "SFR" is the star formation rate, "lifetime" is the lifetime of the WR star in a WR+O binary, which was derived in \citet{Xu2025A&A...704A.218X}. We refer to Appendix~E in \citet{Xu2025A&A...704A.218X} for the details of this method. In order to better reproduce the observed number of the SMC WR stars, we adopted a higher SFR of 0.0821$\msoy$ than the value of 0.05$\msoy$ used in \citet{Xu2025A&A...704A.218X}. With this higher SFR, we re-derived the WR star population from the fiducial population synthesis model in \citet{Xu2025A&A...704A.218X} for comparison purposes (hereafter Simulation2).

In this work, we defined runaway WR stars as the WR stars formed in disrupted binaries, regardless of their space velocities. Binary disruption can take place only if $\Delta M_{\rm dis}$ is lower than the envelope mass of the primary star, which means that the primary star is partially stripped at disruption. We simply assumed that, after disruption, the runaway primary star ejects the remaining envelope within an ignorable timescale.

\begin{figure*}
    \centering
    \includegraphics[width=0.49\linewidth]{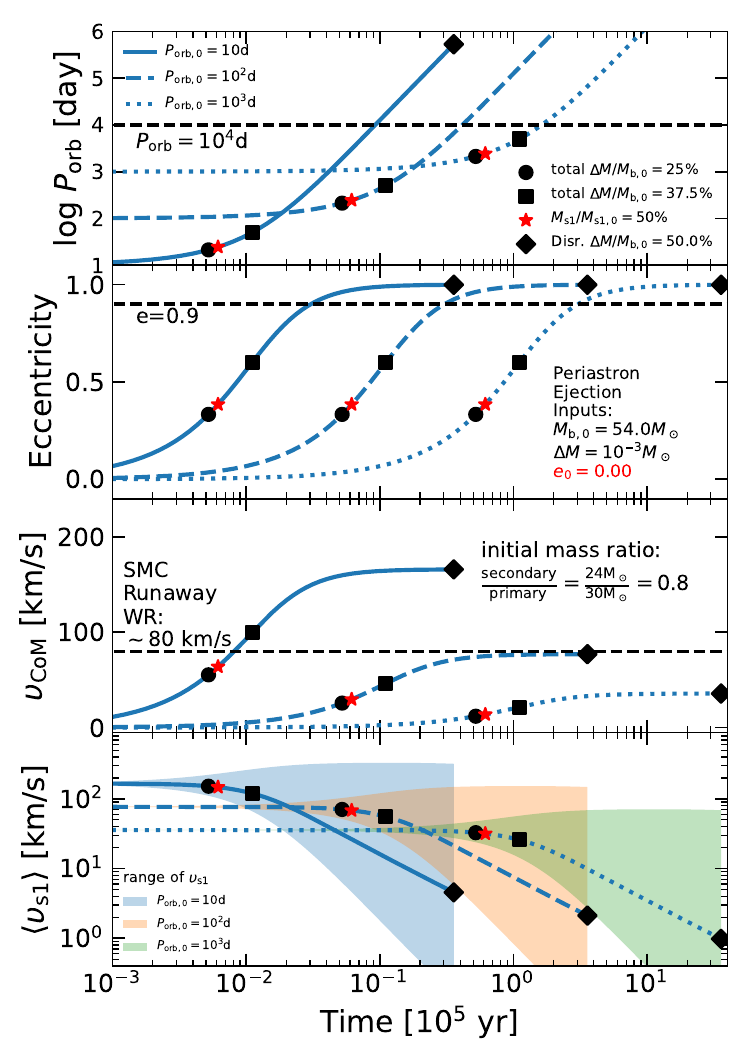}
    \includegraphics[width=0.49\linewidth]{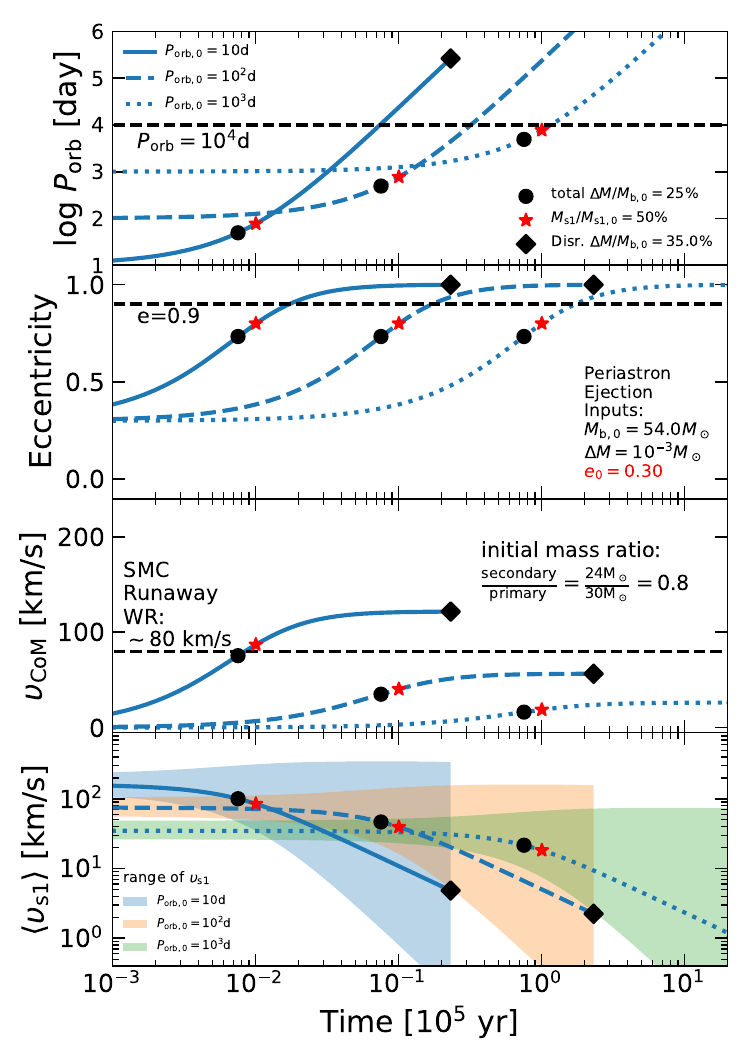}
    \caption{Example models using the mass-ejection-driven orbital evolution, assuming that the primary star ejects a mass of $10^{-3}\mso$ during each periastron passage. The two columns correspond to two initial eccentricities, 0.0 (left) and 0.3 (right). From top to bottom, the panels present the evolution of orbital period, eccentricity, the velocity of the centre of mass, and the average orbital velocity of the mass-losing star. Each panel contains three curves, computed with different initial orbital periods, 10\,d (solid line), 100\,d (dashed line), and 1000\,d (dotted line). The markers correspond to ejected masses of $25\%M_{\rm b,0}$ (circle), $37.5\%\mso$ (square), and $\Delta M_{\rm dis}$ (Eq.\,\ref{dm_per_disr}; diamond). In the $\langle\upsilon_{\rm s1}\rangle$ panel, we use coloured regions to indicate the ranges of the orbital velocities of the mass-losing stars of different initial orbital periods (blue: 10\,d; brown: $10^2$\,d; green: $10^3\,$d), where the upper and lower limits are determined by the velocities at periastron (Eq.\,\ref{eq:v-s1-per}) and apastron (Eq.\,\ref{eq:v-s1-ap}).    
    The red star means that the mass-losing star ejects half of its initial mass. The dashed lines mark $P_{\rm orb}=10^4\,$d (assumed detection limit), $e=0.9$ \citep[the measured eccentricity of WR~140;][]{Thomas2021}, and $\upsilon_{\rm CoM}=80\kms$ \citep[the observed peculiar velocities of SMC~AB1 and AB11;][]{Schootemeijer2024}.}
    \label{fig:example-track}
\end{figure*}

\section{Results\label{sec:results}}

We first present example models of the mass-ejection-driven orbital evolution  to illustrate the main features of the evolution (Sect.\,\ref{sec:example-model}), and we apply our model to WR~140 to infer the properties of its progenitor system (Sect.\,\ref{sec:wr140}). We present our population synthesis predictions for the SMC WR stars in Sect.\,\ref{sec:popsync}.

\subsection{Example models\label{sec:example-model}}

In the classical theory of orbital evolution, mass transfer redistributes the masses and angular momenta of stellar components, accompanied by mass loss, carrying away orbital angular momentum \citep{Tauris2006,Tauris2023pbse.book.....T,Marchant2024ARA&A..62...21M}. This process does not alter the velocity of the centre of mass, and the orbit is expected to stay circularised due to the strong tidal torque for a Roche-lobe filling donor. In contrast, the mass-ejection-driven orbital evolution leads to a rapid increase in the orbital period, accompanied by an efficient increase in eccentricity. Also, the sudden mass ejections impart momentum kicks on the binary system, producing measurable space velocities.

Figure\,\ref{fig:example-track} presents three examples of mass-ejection-driven orbital evolution. The pre-mass-ejection orbital periods were taken to be 10\,d, 100\,d, and 1000\,d. The initial total mass of the binaries was fixed to be $54\mso$  with an initial mass ratio of 0.8 (i.e., $30\mso$ primary + $24\mso$ secondary). 
We assumed that each mass ejection event removes $10^{-3}\mso$, which appears achievable for LBVs \citep{Vink2002A&A...393..543V,Smith2006ApJ...645L..45S,Grafener2012A&A...538A..40G,Smith2014ARA&A..52..487S,Sanyal2015,Campagnolo2018A&A...613A..33C,Grassitelli2021A&A...647A..99G,Cheng2024ApJ...974..270C,Pauli2026}. We considered two initial eccentricities, 0 (left column) and 0.3 (right column). We marked three specific values, which are $\porb=10^4\,$d \citep[a period that cannot be excluded by past $\Delta$RV searches; see Fig. 6 in][]{Schootemeijer2024}, $e=0.9$ \citep[the eccentricity of WR~140;][]{Williams2019MNRAS.488.1282W,Thomas2021}, and $\upsilon_{\rm CoM}=80\kms$ \citep[the peculiar velocity of the runaway WR stars in the SMC;][]{Schootemeijer2024}. 
We computed these models until binary disruption for presentation purpose, while, in a realistic binary evolution model, the termination point of the mass-ejection-driven orbital evolution is determined by the stripping of the hydrogen-rich envelope of the LBV primary star.

Closer binaries evolve faster due to a higher frequency of periastron passages.
Our example models with circular initial orbits take about 0.07--1.21 $\times10^5\,$yr to reach $P_{\rm orb} = 10^4\,$d for $P_{\rm orb,0}$ varying from 10\,d to 1000\,d, and the corresponding mass ejection fraction, $\Delta M/M_{\rm b,0}$, is about 0.49--0.42.
Meanwhile, all models reach $e=0.9$ at $\Delta M/M_{\rm b,0}=0.47$, as the evolution of eccentricity does not depend on orbital period (Eq.\,\ref{eq:eper2}).
Since $\upsilon_{\rm CoM}$ is proportional to the relative velocity at periastron, which remains unchanged during the evolution, a closer initial orbit achieves a higher $\upsilon_{\rm CoM}$ at the same $\Delta M/M_{\rm b,0}$. 
The $\upsilon_{\rm CoM}$ at disruption are $165\kms$, $77\kms$, and $35\kms$, corresponding to initial orbital periods of 10\,d, 100\,d, and 1000\,d, where only the model with $P_{\rm orb,0}=10\,$d reaches close to $80\kms$ when the primary star loses half of its initial mass. The average orbital velocities of the mass-losing stars decrease with the increasing orbital periods, which are below $\sim10\kms$ when reaching $P_{\rm orb}=10^4\,$d. 

Binaries with higher initial eccentricities are more easily disrupted. According to Eq.\,\eqref{dm_per_disr}, it requires a $\Delta M/M_{\rm b,0}$ of 50\% to disrupt an initially circular system, which is only 35\% in the case of  $e_0=0.3$. Consequently, the models with $e_0=0.3$ reach $P_{\rm orb}=10^4\,$d and $e=0.9$ at a much lower $\Delta M/M_{\rm b,0}$. Initially eccentric binaries also have higher space velocities at the same $\Delta M/M_{\rm b,0}$ due to the difference in the relative velocities at periastron (Eq.\,\ref{eq:vper0}). Hence, if the observed runaway WR stars in the SMC were formed through the mass-ejection-driven orbital evolution, the progenitor systems were likely to be very eccentric.

The typical lifetime of an LBV is about a few $10^{4}\,$yr \citep{Humphreys1994PASP..106.1025H,Smith2014ARA&A..52..487S}. The envelope mass of a massive star at the SMC metallicity is about half of the ZAMS mass of the star \citep{Xu2025A&A...704A.218X}. For an initial mass ratio of 0.8, the primary star ejects half of its initial mass at $\Delta M/M_{\rm b,0}\simeq 38.5\%$ (red stars in Fig.\,\ref{fig:example-track}), which occurs within a few percent of the time needed for binary disruption.
Meanwhile, the mass ejected during one periastron passage could be much higher than the value we assumed. A stronger mass ejection at periastron can significantly reduce the evolutionary timescale.  Observationally, there are records for much more eruptive events, like the 1600 AD eruption of P Cygni \citep[$0.1\mso$ ejected;][]{Smith2006ApJ...638.1045S}, and the Great Eruption of $\eta$\,Carinae in the 19th century \citep[$10\mso$ ejected;][]{Smith2003}. Therefore, the proposed mass-ejection-driven orbital evolution does not contradict the lifetime of LBVs. 

\subsection{The progenitor of WR~140\label{sec:wr140}}

\begin{table}[t]
    \centering
    \caption{Observed properties of WR~140. \label{tab:wr140-obs}}
    \begin{tabular}{l c|l}\hline\hline
    \multicolumn{3}{c}{}\\[-2ex]
    \multicolumn{2}{l|}{WR140} & Refs\\\hline
    &\\[-2ex]
    O-type star mass ($M_{\rm O}$)  &  $29.27\mso$ & (1)\\ 
        WR companion ($M_{\rm WR}$) & $10.31\mso$& (1)\\
        orbital period ($P_{\rm orb,obs}$)& 2895\,d& (1)\\
        eccentricity ($e_{\rm obs}$) & 0.8993& (1) \\
        proper motion & $\sim34\pm2\kms$& (2)
        \\\hline
    \end{tabular}
    \tablefoot{(1) \citet{Thomas2021}; (2) \citet{Dzib2009RMxAA..45....3D}.}
    \vskip 0.5cm
    \caption{Definitions of our population synthesis models.}
    \begin{tabular}{l|cccc}
    \hline\hline
         &  Simulation1 & High-$M_{\rm crit}$ & High-$P_{\rm crit}$ & Flat-$f_e$\\
         \hline
        $M_{\rm crit}$ [$\mso$] & 30 & 40 & = & = \\
        $P_{\rm crit}$ [d]& 40 & = & 60 & = \\
        $\eta$ & 0.5 & = & =  & 0\\ \hline
    \end{tabular}
    \tablefoot{"=" means the same value as that in Simulation1. }
    \label{tab:popsync-model}
\end{table}

\begin{figure}[t]
    \centering
    \includegraphics[width=\linewidth]{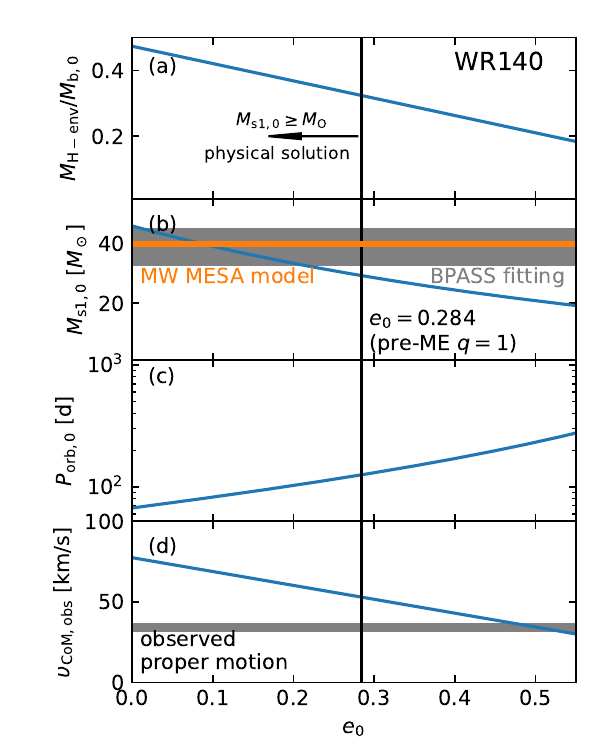}
    \caption{Inferred properties of the progenitor of WR~140 as a function of the pre-mass-ejection eccentricity. Panel\,(a) presents the inferred mass ejection fraction, which is the mass of the hydrogen-rich envelope of the primary star divided by the pre-mass-ejection binary mass. Panel\,(b) presents the inferred pre-mass-ejection mass of the WR progenitor, using the BPASS model \citep{Thomas2021} and the MESA model \citep{Jin2024A&A...690A.135J,Jin2025}, respectively, for comparison. 
    Panel\,(c) presents the pre-mass-ejection orbital period. Panel\,(d) presents the predicted current velocity of the centre of mass, where the grey region indicates the observed proper motion of WR~140 \citep{Dzib2009RMxAA..45....3D}. }
    \label{fig:wr140}
\end{figure}

Since the final orbital parameters of the mass-ejection-driven orbital evolution only depends on the total ejected mass (i.e., the hydrogen-rich envelope of the mass-losing star), it can be used to infer the pre-mass-ejection properties of observed wide and eccentric WR+O binaries. We rewrote Eqs.\,\eqref{eq:aper2}-\eqref{eq:vcomper1} as
\begin{equation}
    \frac{M_{\rm H-env}}{M_{\rm b,0}} = \frac{e_{\rm obs} - e_0}{1+e_{\rm obs}},
    \label{eq:fdm-infer}
\end{equation}
\begin{equation}
    M_{\rm s1,0} = \frac{M_{\rm b,obs}}{1-(M_{\rm H-env}/M_{\rm b,0})} - M_{\rm O},
    \label{eq:Ms1-infer}
\end{equation}
\begin{equation}
    P_{\rm orb,0}^2 = P_{\rm orb,obs}^2 \frac{[(1-e_0) - 2(M_{\rm H-env}/M_{\rm b,0})]^3}{[1-(M_{\rm H-env}/M_{\rm b,0})]^2(1-e_0)^3},
    \label{eq:porb-infer}
\end{equation}
\begin{equation}
    \upsilon_{\rm CoM,obs} =\frac{M_{\rm H-env} M_{\rm O}}{M_{\rm b,0} (M_{\rm b,0}-M_{\rm H-env})}\upsilon_{\rm per,0},
\end{equation}
where the subscript "0" indicates the value before the onset of the mass-ejection-driven orbital evolution, the subscript "obs" corresponds to the observed parameter of WR+O binaries, $M_{\rm H-env}$ is the mass of the hydrogen-rich envelope of the WR progenitor, and $M_{\rm O}$ is the mass of the O star companion. With these equations, the pre-mass-ejection properties can be solved by varying the pre-mass-ejection eccentricity, $e_0$, as a free parameter.

We used the above equations to infer the properties of the progenitor of WR~140, whose observed parameters are summarised in Tab.\,\ref{tab:wr140-obs}. Our results are presented in Fig.\,\ref{fig:wr140}, where $e_0$ is treated as a free parameter, and we infer the other pre-mass-ejection parameters and predict the current space velocity of WR~140, $\upsilon_{\rm CoM,obs}$. During the mass-ejection-driven orbital evolution, the mass of the O star companion was expected to remain unchanged. Hence, physical solutions should meet $M_{\rm s1,0} \geq M_{\rm O}$, which corresponds to the parameter space with $e_0 \leq 0.284$. The mass of the WR progenitor is limited to $30\text{--}45\mso$, which is consistent with the value inferred by using the BPASS model \citep{Thomas2021} and the MESA models computed with Galactic metallicity by \citet{Jin2024A&A...690A.135J} and \cite{Jin2025}\footnote{The model data is available:\newline
\url{https://wwwmpa.mpa-garching.mpg.de/stellgrid/}.}. We describe the method of fitting with the MESA models in Appendix\,\ref{app:mesa-fit}.
While the current orbital period of WR~140 is about 2895\,d, we expect the progenitor system to have an orbital period below $\sim$100\,d. In addition, we predict the current space velocity of WR~140 to be about $50\text{--}75\kms$, which does not contradict the observed proper motion of WR~140 \citep[$\sim$$34\kms$;][]{Dzib2009RMxAA..45....3D}. Therefore, our mass-ejection-driven orbital evolution can self-consistently explain the formation of WR~140-like systems.

\subsection{Population synthesis predictions\label{sec:popsync}}

\subsubsection{Predictions with fixed initial eccentricities\label{sec:popsync-fixed-e0}}

\begin{figure*}[ht!]
    \centering
    \includegraphics[width=0.49\linewidth]{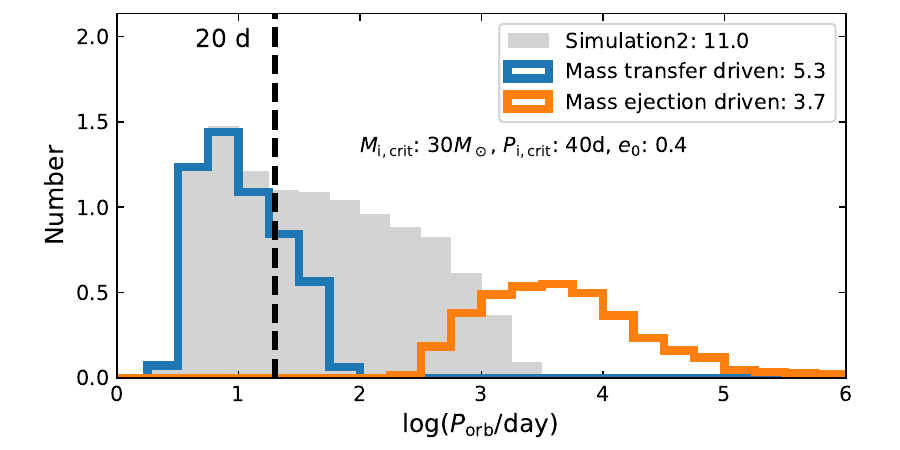}
    \includegraphics[width=0.49\linewidth]{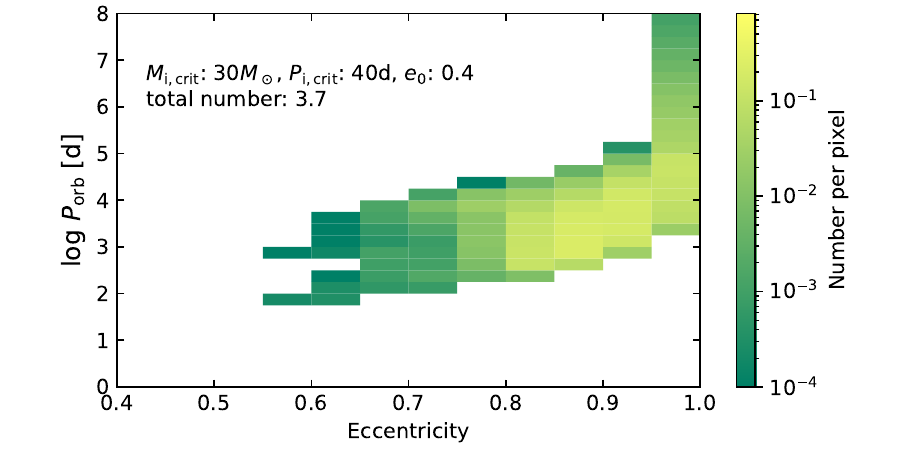}
    \includegraphics[width=0.49\linewidth]{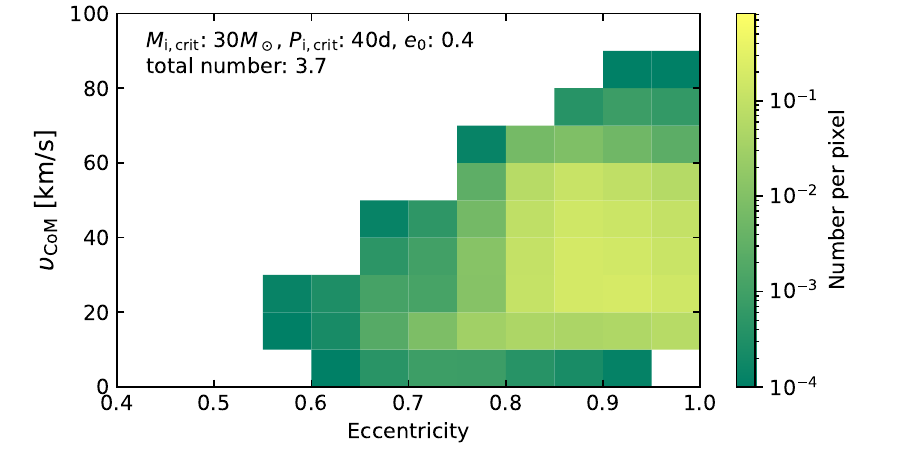}
    \includegraphics[width=0.49\linewidth]{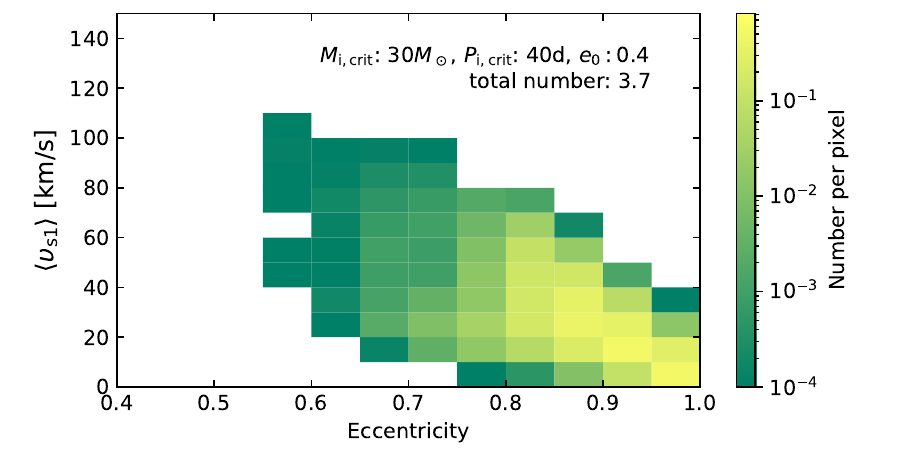}
    \includegraphics[width=0.49\linewidth]{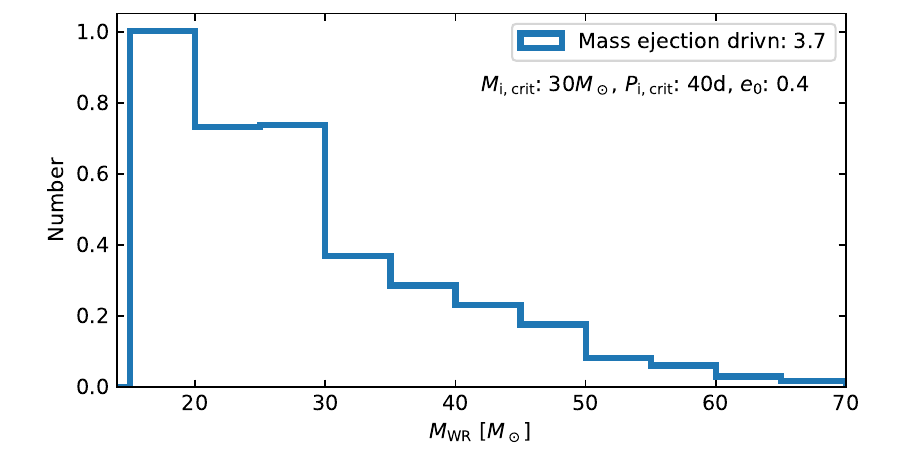}
    \includegraphics[width=0.49\linewidth]{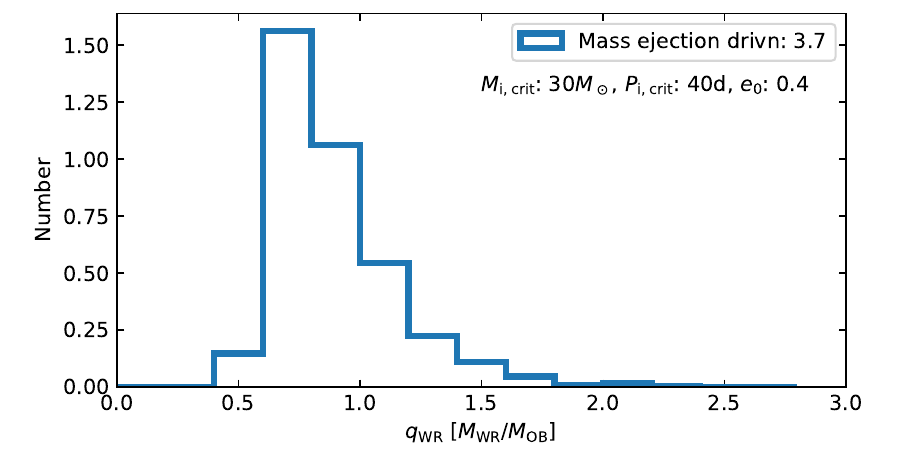}
    \includegraphics[width=0.49\linewidth]{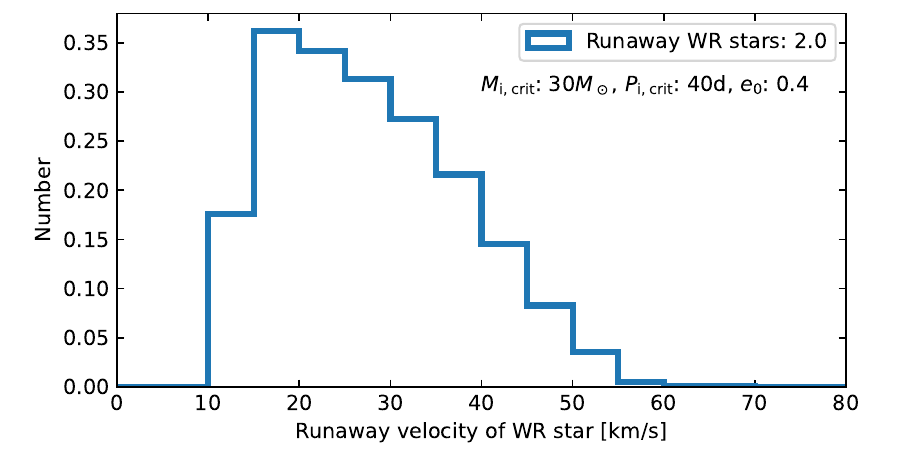}
    \caption{Population synthesis predictions with a fixed initial eccentricity, 0.4. The upper-left panel presents the predicted orbital period distribution, where the systems formed through the mass-transfer- and mass-ejection-driven orbital evolution are plotted in blue and orange, respectively, and the text indicates the predicted numbers. The vertical dashed line marks the longest orbital period (20\,d) of the observed WR+O binaries in the SMC. The prediction by Simulation2, which is the fiducial model in \citet{Xu2025A&A...704A.218X} but with a higher star formation rate, is plotted in grey for comparison.
    The 2D distributions are the mass-ejection WR+O population on the eccentricity-$\logp$ (upper-right), -$\upsilon_{\rm CoM}$ (middle-left) and -$\langle\upsilon_{\rm s1}\rangle$ (middle-right) planes, where the predicted number in each pixel is colour-coded. The third row presents the predicted distributions of WR star masses ($M_{\rm WR}$) and mass ratios (WR star/O star) of the mass-ejection WR+O binaries.
    The bottom panel presents the predicted space velocities of the runaway WR stars.
    }\label{fig:result-popsync-e0-0.4}
\end{figure*}

\begin{figure*}[ht!]
    \centering
    \includegraphics[width=0.49\linewidth]{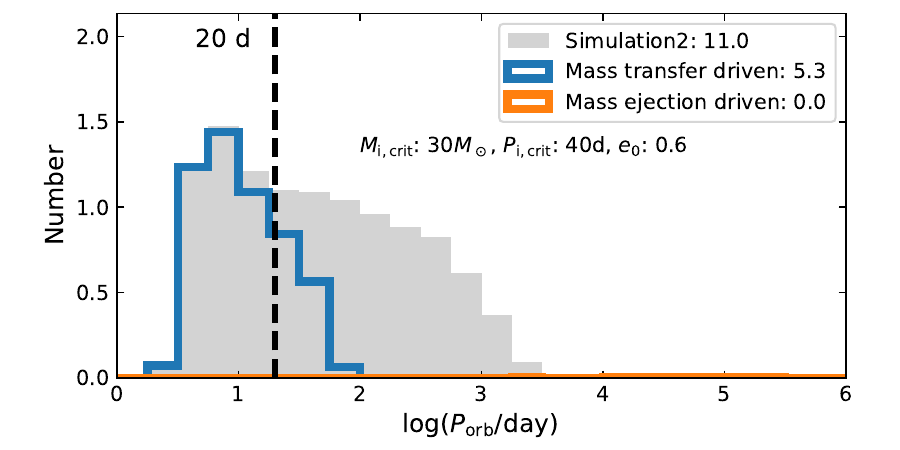}
    \includegraphics[width=0.49\linewidth]{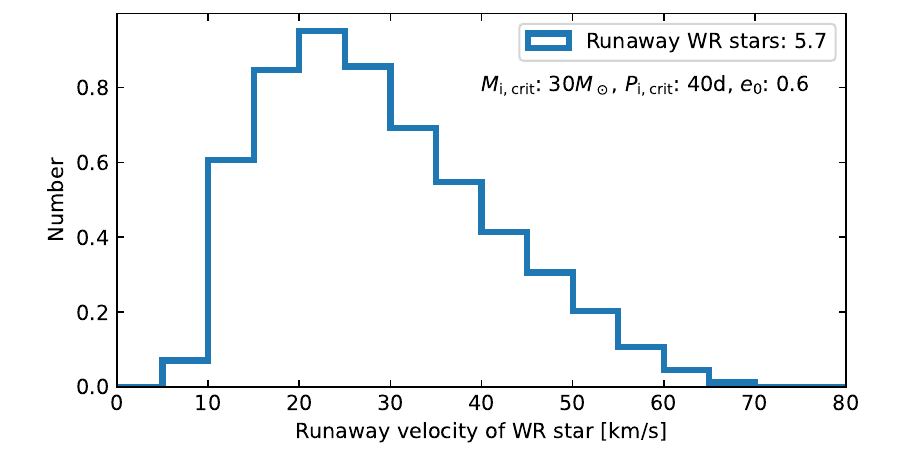}
    \caption{Population synthesis predictions with a fixed initial eccentricity, 0.6. The left panel presents the predicted orbital period distribution, and the right panel presents the predicted space velocities of the runaway WR stars. The colours have the same meaning as in the upper-left and bottom panels of Fig.\,\ref{fig:result-popsync-e0-0.4}.}
    \label{fig:result-popsync-e0-0.6}
\end{figure*}

To more easily interpret our population synthesis predictions from Simulation1, we first present results obtained with fixed initial eccentricities (0.4 and 0.6), together with the corresponding values of $M_{\rm crit}$ and $P_{\rm crit}$ adopted in Simulation1. Results for other initial eccentricities are provided in Appendix\,\ref{app:popsync-fixed-e0-more}. 

Without the mass-ejection-driven orbital evolution, Simulation2 predicts 11 WR+O binaries in the SMC, whose orbital period distribution peaks at $\sim$$10\,$d and decays to $\sim$$1000\,$d, of which 4.3 WR+O binaries are predicted to have orbital periods below the observed longest orbital period, 20\,d, which is consistent with the observed number of close SMC WR+O binaries \citep[4;][]{Shenar2016,Shenar2018,Schootemeijer2018,Schootemeijer2024}. According to our criterion of entering the mass-ejection-driven orbital evolution, 5.3 WR+O binaries are formed from the mass-transfer-driven orbital evolution (hereafter mass-transfer WR+O binaries), whose number rapidly drops towards the long-period regime ($\sim$100\,d), and 5.7 WR stars are formed from the mass-ejection-driven orbital evolution, of which the fractions that remain bound (hereafter mass-ejection WR+O binaries) and get disrupted (runaway WR stars) sensitively depend on the assumed initial eccentricities.

Figure\,\ref{fig:result-popsync-e0-0.4} presents the result with $e_0=0.4$, where 3.7 mass-ejection WR+O binaries are expected, whose orbital periods are widely distributed from about 300\,d to $10^5\,$d with a peak at around 3000\,d. The number drops towards the long-period regime is caused by binary disruption and the effect of the initial orbital period function.
According to Eq.\,\eqref{eq:Pper2}, the orbital period increases less in binaries with higher initial mass ratios, (i.e., higher value for $M_{\rm b,0}$). The shortest orbital period corresponds to an initial orbital period near $P_{\rm crit}$ and an initial mass ratio of near 1, while longer orbital periods can be produced by a wider initial parameter space. Consequently, our obtained distribution drops towards short periods.
The mass-transfer WR+O binaries have a maximal orbital period about 100\,d, which can reach up to 1000\,d for a higher $M_{\rm crit}$ (cf., Sect.\,\ref{sec:param-study} and App.\,\ref{app:param-study}). 

The 2D distributions show that the predicted $P_{\rm orb}$, $\upsilon_{\rm CoM}$, and $\langle\upsilon_{\rm s1}\rangle$ are correlated with eccentricity, which can be explained by Eqs.\,\eqref{eq:e-porb}-\eqref{eq:e-vorb}.
The majority of these long-period ($10^3\text{--}10^4\,$d) binaries are near disruption, featuring eccentricities of 0.8--1.0. The corresponding velocities of the centre of mass are about 20--60$\kms$. Since high-eccentricity systems tend to have wide orbits, more eccentric binaries are associated with lower average orbital velocities for the WR stars, $0\text{--}40\kms$. 

The distribution of the masses of the WR stars in these mass-ejection WR+O binaries peaks at about $15\mso$ and decays towards the high-mass end due to the effect of the initial mass function. The model grid in \citet{Xu2025A&A...704A.218X} is not linearly spaced in initial primary mass, and this grid effect is likely responsible for the flat feature in [$20\mso,\,30\mso$]. The mass ratio ($q_{\rm WR}$: WR star/OB star) distribution peaks at about 0.7. Here $q_{\rm WR}$ can be estimated by 
\begin{equation}
    q_{\rm WR}=\frac{M_{\rm WR}}{M_{\rm OB}}=\frac{M_{\rm WR}/\mi}{M_{\rm OB}/\mi}\approx\frac{0.5}{\qi},
\end{equation}
where $M_{\rm OB}$ is the mass of the OB star companion.
Therefore, the peak of the $q_{\rm WR}$ distribution at 0.7 also corresponds to a $\qi$ of about 0.7. Binaries with $\qi>0.7$ are lacking because most of them merge in \citet{Xu2025A&A...704A.218X} due to the occurrence of inverse mass transfer with a post-MS mass gainer. The number drops towards the high-$q_{\rm WR}$ regime because a high $q_{\rm WR}$ is associated with a low $\qi$, where \citet{Xu2025A&A...704A.218X} expect a large merger fraction \citep[see also][]{Langer2020,Henneco2024A&A...682A.169H}. 
In \citet{Xu2025A&A...704A.218X}, these binaries are assumed to merge during the mass transfer phase, since the combined radiation power of both stars in the systems is not strong enough to expel the material that cannot be accreted by the spun-up main-sequence mass gainer, where the rotation-limited accretion efficiency was adopted \citep[see also][]{Langer2020,Fragos2023ApJS..264...45F}. 
These binaries can avoid merging if the mass transfer phase that causes the mergers is avoided by an LBV stage of the primary star, which our population synthesis cannot assess, as our calculation is limited by the model data of \citet{Xu2025A&A...704A.218X}.

There are about 2.0 runaway WR stars with $e_0=0.4$, most of which have runaway velocities of $\sim20\kms$. The runaway velocities are approximated by the velocities of the centre of mass at disruption (Eq.\,\ref{eq:vcomdis}), which is proportional to $P_{\rm orb,0}^{-1/3}$ (Eq.\,\ref{eq:vper0}). Hence the number drop towards the high-velocity regime is due to the effect of the initial orbital period function.

The lowest eccentricity we predict is about 0.55 with $e_0=0.4$, corresponding to a mass ejection fraction of about $10\%$. We note that the predicted population with eccentricities close to the assumed initial eccentricity should be treated with caution, as we notice that \citet{Xu2025A&A...704A.218X} misclassified a few mass-transferring systems as WR+O binaries, which only takes a small fraction of the predicted population by \citet{Xu2025A&A...704A.218X}, and should not affect the main conclusion of this work. 

By taking $e_0=0.6$, all binaries entering the mass-ejection-driven orbital evolution get disrupted (Fig.\,\ref{fig:result-popsync-e0-0.6}) because of a lower $\Delta M_{\rm dis}$, which can be expressed as a function of $\mi$ and $\qi$, 
\begin{equation}
    \frac{\Delta M_{\rm dis}}{\mi} = \frac{1}{2} (1+\qi)(1-e_0).
\end{equation}
The binaries with higher $\qi$ are harder to disrupt. With $\qi=1$ and $e_0=0.6$, we have $\Delta M_{\rm dis}/{\mi} = 0.4$.
Most of the primary stars lose about half of their initial masses during the mass-ejection-driven orbital evolution (i.e., $M_{\rm H-env}/\mi=0.5$), which exceeds the amount required for disruption. The space velocities of runaway WR stars have a similar distribution as the $e_0=0.4$ case.

\subsubsection{Predictions with the initial eccentricity function}

\begin{figure*}[t]
   
    \centering
    \includegraphics[width=0.49\linewidth]{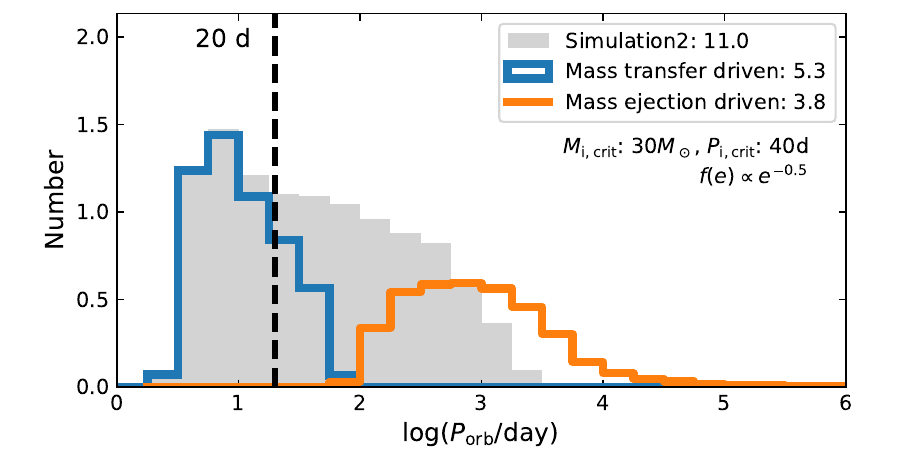}
    \includegraphics[width=0.49\linewidth]{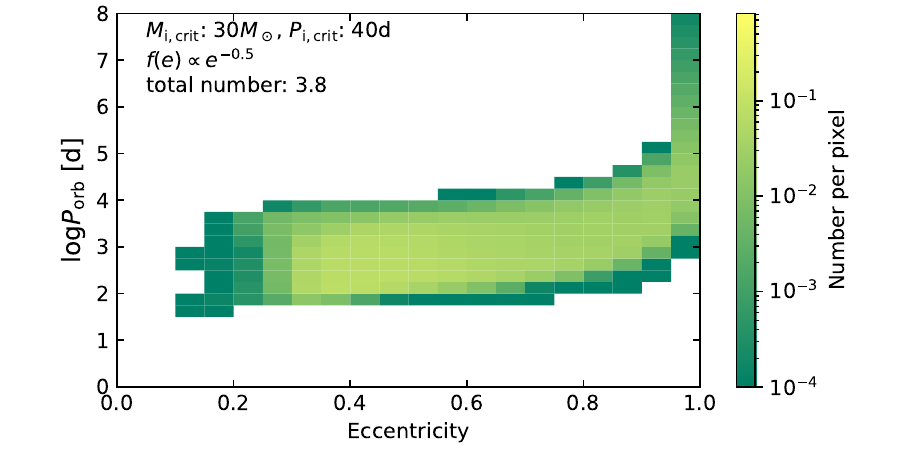}
    \includegraphics[width=0.49\linewidth]{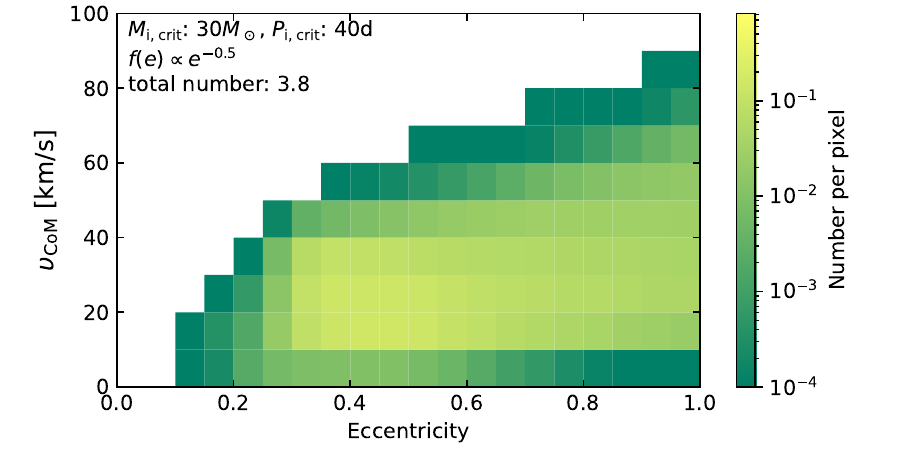}
    \includegraphics[width=0.49\linewidth]{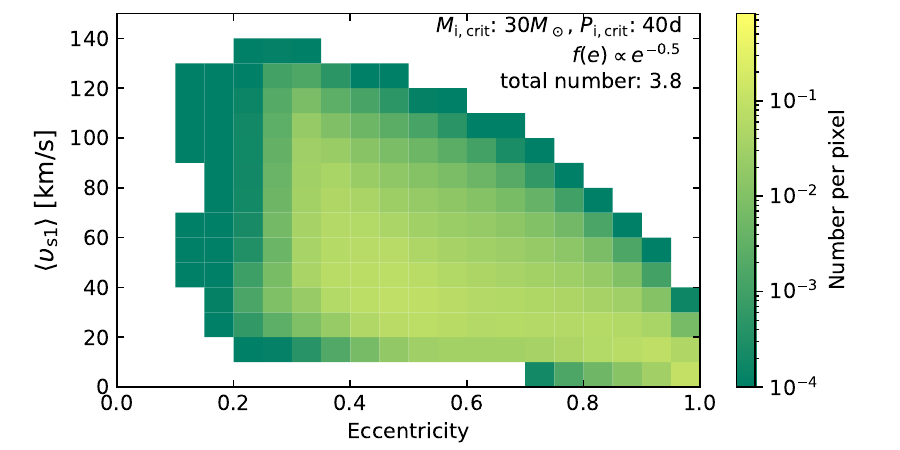}
    \includegraphics[width=0.49\linewidth]{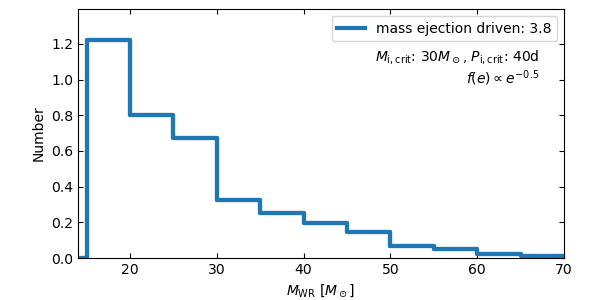}
    \includegraphics[width=0.49\linewidth]{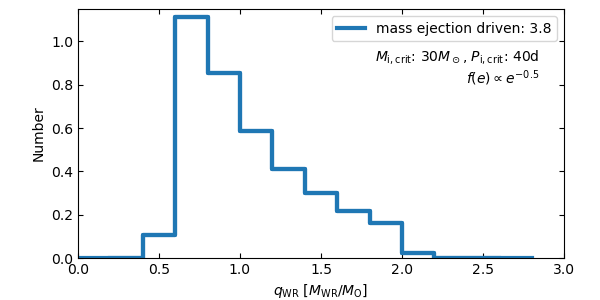}
    \includegraphics[width=0.49\linewidth]{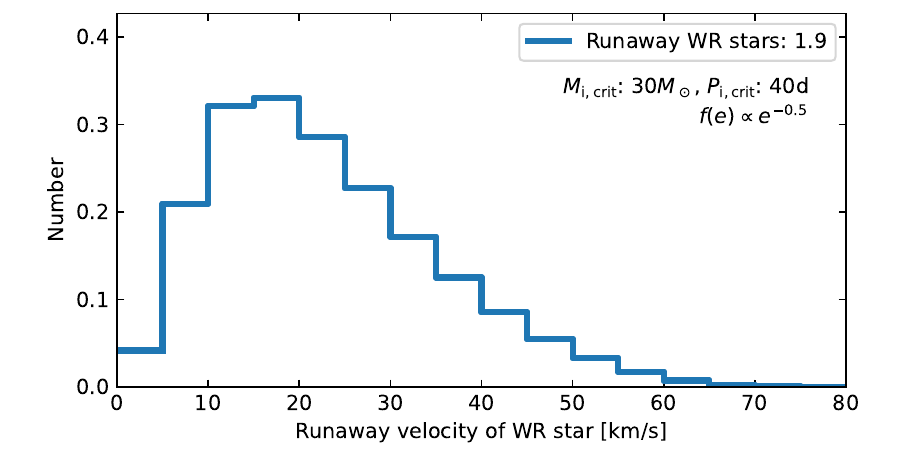}
    \caption{Population synthesis predictions with the initial eccentricity function in \citet{Sana2012} (i.e., Simulation1 in Tab.\,\ref{tab:popsync-model}). The colours have the same meanings as Fig.\,\ref{fig:result-popsync-e0-0.4}.\label{fig:popsync-fe}}
\end{figure*}

The properties of the synthetic population by Simulation1 are presented in Fig.\,\ref{fig:popsync-fe}, which shows similar features as the predicted population with  $e_0=0.4$ (Fig.\,\ref{fig:result-popsync-e0-0.4}). With $\eta=0.5$, about 63\% of pre-mass-ejection binaries have eccentricities below 0.4 according to Eq.\,\eqref{eq:fe-fraction}. Consequently, the predicted orbital periods of the mass-ejection WR+O binaries, peaking at around 562\,d (2.75 in log), are generally shorter than in the $e_0=0.4$ case (Fig.\,\ref{fig:result-popsync-e0-0.4}), and the expected average orbital velocities ($\langle\upsilon_{\rm s1}\rangle$:\,$20\text{--}100\kms$), higher than the $e_0=0.4$ case.
For the same reason, the predicted eccentricities are widely distributed from about 0.4 to 1. The predicted distributions of $M_{\rm WR}$ and $q_{\rm WR}$ show the same peak value as the $e_0=0.4$ case, which are $\sim15\mso$ and $\sim0.7$ respectively. The predicted $\upsilon_{\rm CoM}$ covers the peculiar velocities of SMC~AB1 and SMC~AB11 \citep[$\sim80\kms$;][]{Schootemeijer2024}, but the high-velocity population only takes a small fraction of the whole population. In addition, we predict 1.9 runaway WR stars, whose runaway velocity distribution peaks at 10--20$\kms$ and decays to about 60$\kms$.

\subsubsection{Parameter study\label{sec:param-study}}

By varying the values of $M_{\rm crit}$, $P_{\rm crit}$, and $\eta$, we constructed three population synthesis models, besides Simulation1, whose properties are summarised in Tab.\,\ref{tab:popsync-model}. 
Table\,\ref{tab:param-study} presents the predictions by different population synthesis models. We provide the predicted distribution functions in Appendix\,\ref{app:param-study}.
The main differences lie in the predicted numbers and predicted orbital period distributions. 

A higher $M_{\rm crit}$ can largely reduce the predicted number of mass-ejection WR+O binaries due to the effect of the initial mass function. The High-$M_{\rm crit}$ model predicts 8.1 mass-transfer WR+O binaries (Simulation1: 5.3), of which a considerable fraction have orbital periods above $20\,$d with a maximal orbital period of 1000\,d. This long-period tail does not show up in Simulation1 because all WR star progenitors in initially wide binaries have the LBV phase with $M_{\rm crit}=30\mso$. There are 2.9 WR star systems formed through the mass-ejection-driven orbital evolution, of which 2.0 remain bound, and 0.9 get disrupted, and the peak orbital period of the mass-ejection WR+O binaries is similar to that predicted by Simulation1.

Changing $P_{\rm crit}$ from 40\,d to 60\,d does not significantly alter the predicted numbers. The High-$P_{\rm crit}$ model predicts 6.1 mass-transfer WR+O binaries, 3.2 mass-ejection WR+O binaries, and 1.7 runaway WR star. These predicted numbers are similar to those in Simulation1. As expected, with a higher $P_{\rm crit}$, the maximal orbital period of the mass-transfer WR+O binaries reaches 100\,d (fiducial: 56\,d), and the peak orbital period of the mass-ejection WR+O binaries also becomes slightly longer (from 562\,d to 750\,d).

A flat initial eccentricity distribution leads to fewer low-eccentricity pre-mass-ejection binaries than in Simulation1. For example, 40\% binaries have initial eccentricities below 0.4 in the Flat-$f_e$ model, which is 63\% in Simulation1. Hence, more runaway WR stars can be expected (Flat-$f_e$: 3.2; Simulation1: 1.9). Since binary disruption becomes generally easier than in Simulation1, the peak orbital period of the mass-ejection WR+O binaries is shifted to 1000\,d (Flat-$f_e$) from 562\,d (Simulation1).

\begin{table*}[t]
    \centering
    \caption{Prediction from our different population synthesis models.}
    \begin{tabular}{c|ccc|cccc}\hline\hline
                 &&&&&&&\\[-2.ex]
                 &\multicolumn{3}{c|}{Predicted numbers of WR star systems}&\multicolumn{4}{c}{Predicted orbital period distributions}\\
         &  mass transfer & mass ejection & runaway & $P_{\rm MT, max}$ [d] & (log) &$P_{\rm ME,peak}$ [d] & (log)\\[0.3ex]\hline
             &&&&&&&\\[-2.ex]
    Simulation1 &  5.3& 3.8& 1.9& 56&  (1.75) & $562_{-246}^{+438}$ & ($2.75_{-0.25}^{+0.25}$) \\
    &&&&&&&\\[-2.ex]
    High-$M_{\rm crit}$ & 8.1& 2.0& 0.9& 1000& (3.0) & $422_{-244}^{+578}$ &  ($2.625_{-0.375}^{+0.375}$)\\
    &&&&&&&\\[-2.ex]
    High-$P_{\rm crit}$ & 6.1& 3.2& 1.7& 100& (2.0) & $750_{-434}^{+1028}$ &  ($2.875_{-0.375}^{+0.375}$)\\
    &&&&&&&\\[-2.ex]
    Flat-$f_e$ & 5.3 & 2.5 & 3.2 & 56 & (1.75) & $1000_{-438}^{+778}$&($3_{-0.25}^{+0.25}$)\\[0.5ex]\hline
    \end{tabular}
    \tablefoot{See Tab.\,\ref{tab:popsync-model} for the definitions of the population synthesis models. The 2nd--4th columns are the predicted numbers of mass-transfer WR+O binaries, mass-ejection WR+O binaries, and runaway WR stars, respectively.
    The 5th--8th columns describe the predicted orbital period distributions of mass-ejection/mass-transfer WR+O binaries by using two parameters, which are the maximal orbital period of mass-transfer WR+O binaries, $P_{\rm MT,max}$, and the peak orbital period of the distributions of mass-ejection WR+O binaries, $P_{\rm ME,peak}$ (see Appendix\,\ref{app:param-study} for the predicted distribution functions).}
    \label{tab:param-study}
\end{table*}

\subsection{Second-born Wolf-Rayet stars\label{sec:2nd-bornWR}}

A detailed population synthesis calculation which includes WR stars formed by the initial secondary stars (second-born WR stars) is beyond the scope of this paper. In this subsection, we used simplified assumptions to estimate the population of the second-born WR stars based on the synthetic population of the first-born WR stars in Simulation1.

All WR stars are assumed to be BH progenitors, and we take the mass of the WR star at core helium depletion as the mass of the resulting BH. Consequently, the resulting BH+O binaries have the same orbital properties as their direct progenitors (i.e., the WR+O binaries). Consistent with Simulation1, we apply the criterion described in Sect.\,\ref{sec:popsync} with $(M_{\rm crit},\,P_{\rm crit})= (30\mso,\,40\,\text{d})$ to determine whether the orbital evolution of a BH+O binary model is also driven by LBV mass ejections. If so, we assume that the O star loses half of its mass in the process. \citet{Xu2025A&A...704A.218X} define a WR star as a core helium-burning star with a luminosity above $10^{5.6}\lso$, which corresponds to a minimum WR star mass of about $14\mso$ and a zero-age main-sequence mass of about $30\mso$\footnote{Mass transfer during the main-sequence phase of the donor star produces lighter helium stars  \citep{Sen2022,Schurmann2024A&A...690A.282S}, whose parameter space however is negligible in this study.}. Hence, we only considered the evolution of BH+O binaries with O star companions more massive than $30\mso$. To estimate the number of the WR stars, we calculated their statistical weight (Eq.\,\ref{eq:weight}) with a typical lifetime, 0.4\,Myr, for all second-born WR stars.

Our results are summarised in Tab.\,\ref{tab:2nd-bornWR}. They suggest that second-born WR stars may comprise about 1/3rd of the WR star population. Their number is similar (within a factor 2) to the number
of WR+OB systems because of their similar lifetimes, which are the lifetime of the WR stars (\cite{vdH2017MNRAS.471.4256V}, and Chapter 3 in \cite{Xu2024thesis}). While we find that 
about two third of the second-born WR stars would be born from classical mass transfer (to their
BH companions), about one third is massive enough (cf., the $q_{\rm WR}$ distribution in Fig.\,\ref{fig:popsync-fe}) to evolve through the LBV mass ejection channel. In those, the eccentricities
are already large due to the first LBV driven interaction phase, such that their disruption rate
is higher ($\sim$50\%) than that of the first LBV driven phase ($\sim$33\%).

In our population synthesis calculation, we normalised the predicted number of the first-born WR stars to the observed number (11), which corresponds to a constant star formation rate of 0.0821$\msoy$. After including the second-born WR stars, we find that the observed number can be reproduced by a star formation rate of 0.055$\msoy$, which is well consistent with the value, $0.05\mso$, adopted in \citet{Xu2025A&A...704A.218X}.

\begin{table}[t]
    \centering
        \caption{Predicted fractions (number in each subclass divided by the total number of WR stars) of first- and second-born WR stars in the SMC.}
    \begin{tabular}{r|l}\hline\hline
    \multicolumn{2}{l}{First-born WR star: 67.2\%}\\[0.1 ex]\hline
       mass-transfer WR+O  & 32.4\% \\[0.1 ex]
       mass-ejection WR+O  & 23.2\%\\[0.1 ex]
        runaway WR & 11.6\%\\[0.1 ex]\hline
        \multicolumn{2}{l}{Second-born WR star: 32.8\%}\\[0.1 ex]\hline
        mass-transfer BH+WR & 22.6\%\\[0.1 ex]
        mass-ejection BH+WR & 4.8\%\\[0.1 ex]
        runaway WR & 5.4\%\\\hline
    \end{tabular}
    \tablefoot{The synthetic population of the first-born WR stars is taken from Simulation1, and the population of the second-born WR stars is derived from it with simplified assumptions (see Sect.\,\ref{sec:2nd-bornWR}).}
    \label{tab:2nd-bornWR}
\end{table}

\section{Discussion\label{sec:discussion}}

\subsection{Caveats of our model assumptions\label{sec:caveats}}

We have ignored the effects of steady stellar/LBV winds in the above calculations. The mass loss through steady winds is an adiabatic process for orbital evolution. Hence, we expect it to slowly widen the orbit but preserve the eccentricity. Our population synthesis calculations were performed with the SMC metallicity ($\sim$1/6.5th of the Solar value), where stellar winds are generally weak \citep{Mokiem2007}, and the corresponding orbital widening prior to the LBV stage is ignorable. However, as we will show in Sect.\,\ref{gaiabh}, strong wind mass loss can occur during the main-sequence evolution of Galactic massive stars, which affects our estimation of the properties of the progenitor of WR\,140. Firstly, a strong wind mass loss allows the initial primary to be less massive than its companion at the onset of the LBV stage, while we only considered the solutions where $M_{\rm s1,0}\geq M_\text{O}$ in Fig.\,\ref{fig:wr140}. Secondly, binaries with smaller initial orbital periods could enter our LBV scenario, due to the orbital widening by stellar wind. The steady winds during the LBV stage diminish the orbital widening and eccentricity pumping related to the mass ejection at periastron. Hence, the proposed mass-ejection-driven orbital evolution is significant if the amount of mass ejected at periastron is larger than that lost through steady LBV winds.

Different from the treatment in \citet{Sepinsky2007}, who investigated the orbital evolution of  impulsive mass transfer to the companion star, we assume the ejected material does not interact with the companion star and directly leaves the system.  The companion star could capture a fraction of the ejected material if the velocity of this material is lower than the orbital velocity of the companion star \citep[e.g.,][]{Sen2021}, where the system loses less orbital angular momentum. Our work focuses on wide binaries, where the orbital velocities of the companions are generally below $100\kms$, while  the ejected material from a LBV star can be much faster than this value \citep{Smith2014ARA&A..52..487S}.  Therefore, we consider the wind capture ignorable.

In non-interacting wides binaries, the effects of tides can be safely ignored. However, with increasing Roche-lobe filling factor, tides become more and more significant. In App.\,\ref{app:tides}, we show that the efficiency of the tidal circularisation with an inflated massive star is small compared with the eccentricity pumping related to the mass ejection at periastron. 

Our criterion for the occurrence of mass-ejection-driven orbital evolution uses two fixed parameters, a critical mass $M_{\rm crit}$ and a critical orbital period $P_{\rm crit}$, where the first one guarantees that an inflated envelope is formed, and the second ensures that there is the space for it in the binary system. In Simulation1, we chose the critical mass as $30\mso$ according to the luminosities of observed LBVs. In particular, there is only one confirmed LBV star in the SMC, R\,40, whose luminosity is consistent with the single star model of $30\mso$ \citep{Kalari2018A&A...618A..17K}. Stellar evolution models suggest that envelope inflation at SMC metallicity becomes significant from $40\mso$, with smaller values more appropriate for higher metallicities  \citep{Sanyal2017A&A...597A..71S,Pauli2026}. With $M_{\rm crit} = 40\mso$, some WR star progenitors do not evolve through an LBV stage, which results in WR+O binaries with period above 100\,d formed through mass transfer (cf., the High-$M_{\rm crit}$ model in Fig.\,\ref{fig:popsync-other-models}).

The near-vertical part of the Humphreys-Davidson limit in the HR diagram can be parametrized as $R\sim M^{1/6}$. With that, the orbital period of circular binaries in which the primary star fills its Roche volume at the Humphreys-Davidson limit 
is $P_{\rm HD} \sim f(q)\, [M_{\rm 1}^{0.5}/(M_{\rm 1}+M_{\rm 2})]^{0.5}$, with $f(q)=[0.6\,q^{2/3}+\ln(1+q^{1/3})/(0.49\,q^{2/3})]^{3/2}$. This period is close to $P_{\rm crit}$, and varies only weekly with mass ($P\sim M_1^{-1/4}$, assuming $L\sim M^{1.5}$) and mass ratio ($P\sim q^{-0.07}$, for $q>0.5$). Since lines of constant inflation factors in single star models run close to and parallel to the Humphreys-Davidson limit (cf., Fig.\,1 of \cite{Pauli2026}), our assumption of a constant critical period roughly captures those binaries where the primary star is inflated enough to be considered an LBV. 

\subsection{The number of long-period Wolf-Rayet star binaries\label{sec:dis-compare-obs}}

Our population synthesis predictions show that our mass ejection scenario transforms the period- and eccentricity distributions of the long-period WR binary population. Our Simulation1 predicts 5.3 close WR+O binaries \citep[4 observed in the SMC;][]{Shenar2016,Shenar2018,Schootemeijer2018,Schootemeijer2024}, 3.8 wide WR+O binaries, most of which may not be ruled out by previous radial-velocity searches and exhibit measurable space velocities, and 1.9 runaway WR stars \citep[7 single WR stars observed in the SMC, including 2 runaway stars;][]{Schootemeijer2024}.  The space velocities of the observed runaway WR stars imply that their progenitors may have had relatively high eccentricities and short orbital periods, or the past mass ejections may have generated additional momentum kicks. For the predicted wide WR+O binaries, it might be difficult to detect their binarity, but not impossible. If such systems were truly missing in the SMC, perhaps wide SMC O star binaries are more eccentric than we assumed, and more often disrupted. 
For Simulation2, the synthetic WR+O binaries contains a large fraction of binaries with orbital period between 20\,d and 1\,yr, whose binarity would be identified by \citet{Schootemeijer2024} if they existed. 
Besides potential massive companions suggested by our results, companions below $\sim 5\mso$ to these apparently single SMC WR stars are also not excluded  \citep{Olejak2025arXiv251110728O}.
In addition, \citet{Schootemeijer2024} also report the absence of the He~I~4471~$\mathring{\rm A}$ line and no significant correlation between successive observations and the radial-velocity differences between them, whose implications for the binarity of the SMC WR stars are beyond the scope of this work. 

We note that the above numbers focus on the LBV stage of the primary stars in wide O star binaries. As discussed in Sect.\,\ref{sec:2nd-bornWR}, the secondary stars might also evolve through an LBV stage, triggering a second mass-ejection-driven orbital evolution and forming second-born WR stars, whose number might be about half that of the first-born WR stars. Some apparently single SMC WR stars may also originate from the self-stripping of single stars \citep{Shenar2020}. Assuming 70\% of the SMC O stars are in binary systems, and half of the single O stars are massive enough to reach their Eddington limit and self-strip, about 15\% of the SMC WR stars might be formed through self-stripping. Therefore, our population synthesis calculations might only cover about half of the SMC WR star population.

Our mass-ejection-driven orbital evolution model predicts that the binary fractions roughly remain unchanged from O stars to LBVs, but decrease significantly for WR stars, which is consistent with current observations.
The binary fractions of Galactic O stars and LBVs are observed to be about 70\% \citep{Sana2012} and $62^{+38}_{-24}\%$ \citep[bias corrected binary fraction;][]{Mahy2022A&A...657A...4M}, but the binary fraction of WR stars significantly drops to about 36\% (4/11) according to the SMC sample, or about 40\% according to the LMC sample \citep[107 WN sars, of which  17 confirmed and
22 candidate binary systems;][]{Hainich2014A&A...565A..27H}, suggesting that some of the single WR stars may be WR~140-like systems or runaway stars.

\subsection{Implication for the orbital period distributions of Galactic WNE and WC star binaries}

Carbon-rich WR stars are suggested to evolve from early-type nitrogen-rich WR (WNE) stars through wind mass loss  \citep{Langer1987A&A...171L...1L,Langer2012}. As a consequence, a WNE binary evolves into a WC binary, with its orbital evolution driven by wind mass loss. However, the observed orbital period distributions of Galactic WN and WC binaries seem inconsistent with the above theoretical expectation, peaking at 1--10\,d for WN binaries, but at about 5000\,d for WC binaries \citep{Dsilva2023A&A...674A..88D}. 
The limited sample size might cause this potential tension \citep{Deshmukh2024A&A...692A.109D}. More observational data is required to verify whether the orbital period distributions of WNE and WC binaries are different or not.

If the orbital period distributions of WNE and WC binaries are truly in tension with each other, the mass-ejection-driven orbital evolution proposed in this work may offer a plausible explanation, given that we have shown one of these long-period WC binaries, WR~140, could result from such evolution (cf., Sect.\,\ref{sec:wr140}). 
More massive helium stars spend more of their lifetime as WC stars \citep{Aguilera-Dena2022A&A...661A..60A}, and their progenitor stars have more chance to reach the Eddington limit, triggering the LBV mass ejection and resulting in the long-period WC binaries.
The observed WNE binaries are more likely to originate from less massive systems, since less massive helium stars remain as WN stars during the lifetime  \citep{Aguilera-Dena2022A&A...661A..60A}, and they are preferred by the initial mass function. Most of the progenitor stars are not massive enough to reach the Eddington limit, and hence mass transfer drives the orbital evolution, resulting in an orbital period distribution for WN binaries similar to O star binaries, as found by \citet{Dsilva2023A&A...674A..88D}.

\subsection{Implication for merging binary black holes\label{sec:merging-bbh}}

\begin{figure*}[!t]
    \centering
    \includegraphics[width=\linewidth]{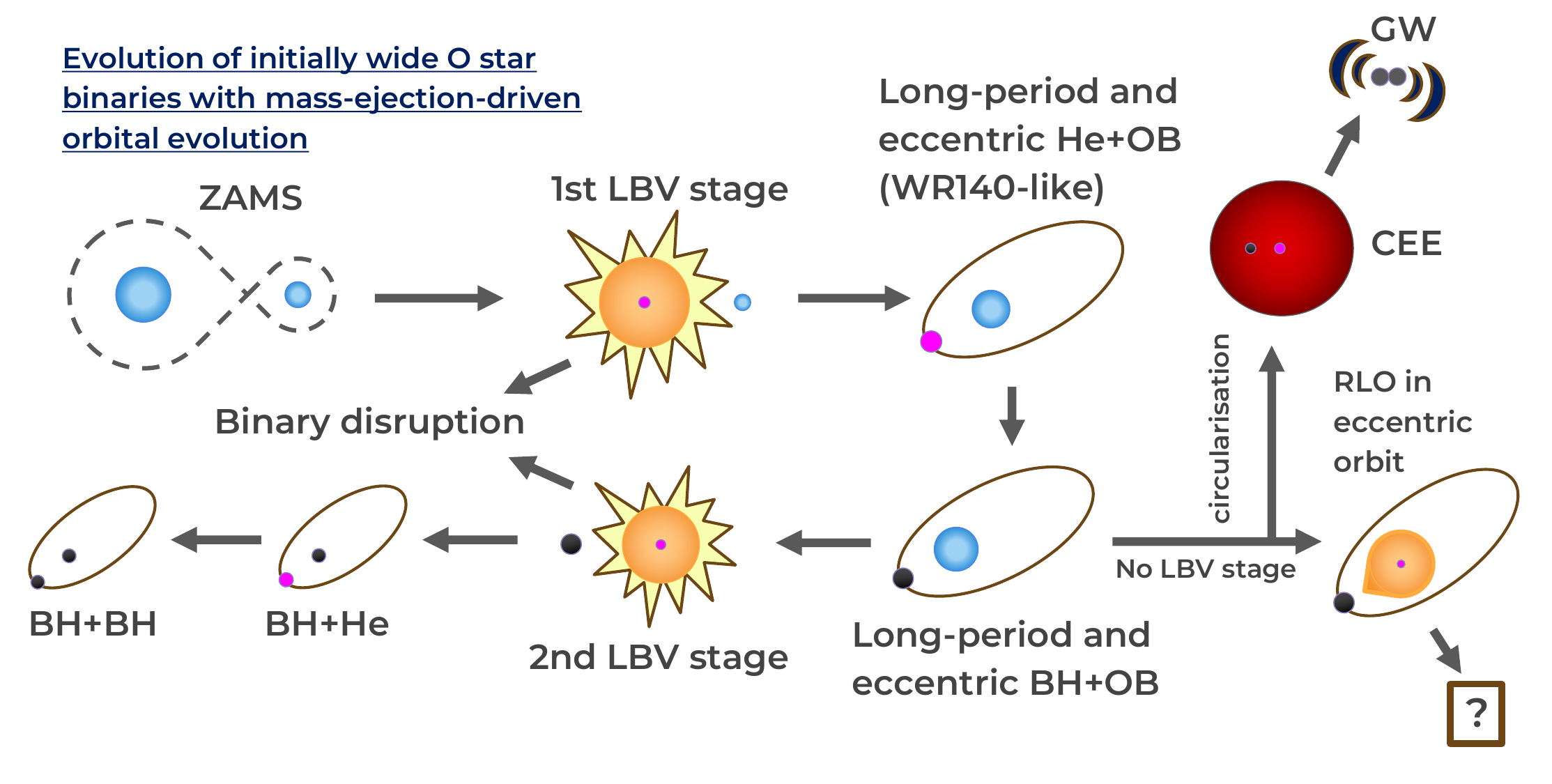}
    \caption{Schematic evolutionary sequence for initially wide O star binaries with the mass-ejection-driven orbital evolution. The abbreviations have the same meaning as Fig.\,\ref{fig:schematic}. \label{fig:schematic-LBV}}
\end{figure*}

As indicated in Fig.\,\ref{fig:schematic}, the lack of long-period WR binaries includes those expected to enter common envelope evolution, which may have strong implications for the formation of merging BBHs. If the O stars in wide binaries disappear because
they merge with their companions during the first mass transfer phase, they cannot form BBHs.

If the mass-ejection-driven orbital evolution is responsible for the lack of observed long-period WR binaries, merging BBHs can hardly be formed in wide binaries either. Figure\,\ref{fig:schematic-LBV} shows a schematic plot of the evolution of initially wide O star binaries with the mass-ejection-driven orbital evolution. It suggests that the long-period WR binaries appear as single stars, or that most of wide O star binaries are disrupted by the LBV mass ejection. 
Then, if the binary can survive the LBV stage, the existence of long-period and eccentric BH+O binaries can be expected, if the WR stars can successfully collapse into BHs. Different from the prediction by \citet{Xu2025A&A...704A.218X} that most of the OB star companions to BHs are fast-rotating, we expect the O stars to remain rotating slowly. 
Then the expansion of the companion star triggers the mass transfer in eccentric binaries \citep{Sepinsky2007,Sepinsky2009ApJ...702.1387S,Rocha_2025,Parkosidis2025arXiv250905243P} or a second episode of mass-ejection-driven orbital evolution. 
The formation of merging BBHs is possible only if the tide is strong enough to circularise the orbit, and the companion star can avoid the LBV phase. Then the future evolution of the system can be described by the mass transfer through Roche-lobe overflow, where merging BBHs are formed through either the CEE channel \citep{Belczynski2016} or the stable mass transfer channel \citep{Marchant2021,Klencki2025arXiv250508860K,Xu2025arXiv251220054X,Briel2026arXiv260203629B}. 
In addition, for eccentric low-mass systems, a similar mechanism, {\em gazing envelope evolution}, has been proposed by \citet{Kashi2018MNRAS.480.3195K}.

\subsection{Implication for Gaia black hole systems} \label{gaiabh}

The recently detected Gaia BH systems contain a low-mass star orbiting a massive BH in a wide and eccentric orbit (see Tab.\,\ref{tab:GaiaBH}). It has been suggested that they have been formed by dynamical interaction in star clusters \citep{Rastello2023MNRAS.526..740R} or in triple star systems \citep{Naoz2025ApJ...992L..12N}. Evolutionary channels in isolated binary systems have also been identified, with some scenarios involving BH kicks \citep{Kotko2024MNRAS.535.3577K,Deng2024ApJ...977...95D}, and others relying on the avoidance of mass transfer from the BH progenitors to the low-mass companions \citep{Gilkis2024MNRAS.535L..44G,Olejak2025arXiv251110728O}. The latter scenarios usually have difficulties producing the eccentricities of Gaia BH1 and BH2, while Gaia BH3 can be explained by wide non-interacting binaries \citep{Iorio2024A&A...690A.144I}. The mass-ejection-driven orbital evolution might provide a mechanism to pump eccentricities in Gaia BH progenitor systems. In the following, we investigate whether our model can provide a formation scenario for the systems Gaia BH1 and BH2, to estimate the properties of their progenitor systems. 

Different to the SMC models, stellar wind mass loss may be important in Galactic massive binaries.  Assuming that the Gaia BHs formed through direct collapse, the initial mass of the progenitor WR star can be estimated by $f_{\rm WR,wind}\,M_{\rm BH}$, where $M_{\rm BH}$ is the BH mass, and the $f_{\rm WR,wind}$ factor captures the effect of WR star winds. We adopted $f_{\rm WR,wind}=2$, implying that a $20\mso$ WR star would lose half of its mass through winds, and produces a $10\mso$ BH. Since the Gaia BHs and their progenitor stars are more massive than their low-mass companions, the binary masses can be approximated by the primary masses\footnote{$M_{\rm b,0}\simeq M_{\rm s1,0}$ in Eq.\,\eqref{eq:fdm-infer}, $M_{\rm b,obs}\simeq M_{\rm BH}$ in Eq.\,\eqref{eq:Ms1-infer}, and the companion mass term, $M_{\rm O}$, in Eq.\,\eqref{eq:Ms1-infer} can also be ignored.}. Then, according to Eqs.\,\eqref{eq:fdm-infer} and \eqref{eq:Ms1-infer}, the ejected mass required to produce the observed eccentricities, $M_{\rm ej,per}$, can be estimated by 
\begin{equation}
    M_{\rm ej,per} \simeq f_{\rm WR,wind}\left[\frac{M_{\rm BH}}{1-(e_{\rm obs}-e_0)/(1+e_{\rm obs}) } - M_{\rm BH}\right].
\end{equation}
For $e_{\rm obs}=0.5$, $e_0 = 0$ (i.e., a circular pre-mass-ejection orbit), and $M_{\rm BH}=10\mso$, our model requires the progenitors of Gaia BH1 and BH2 to eject $10\mso$ during the LBV phase. Based on the MESA models computed by \citet{Jin2025}, this could be achieved from a star with an initial mass of about 50$\mso$, which has lost about $20\mso$ through a wind during its main-sequence evolution, before entering its LBV phase with a 10$\mso$ hydrogen-rich envelope.  

The required pre-LBV orbital separation can be estimated by using Eq.\,\eqref{eq:porb-infer} with Kepler’s third law, which is
\begin{equation}
    \frac{a_{\rm i}^3}{a_{\rm f}^3} = \frac{M_{\rm ej,per}+f_{\rm WR,wind}M_{\rm BH}}{f_{\rm WR,wind}M_{\rm BH}}\frac{[(1-e_0) - 2(M_{\rm H-env}/M_{\rm b,0})]^3}{[1-(M_{\rm H-env}/M_{\rm b,0})]^2(1-e_0)^3},
\end{equation}
where $a_{\rm i}$ and $a_{\rm f}$ are the semi-major axes before and immediately after the LBV phase, respectively. Assuming again $e_0=0$, $a_{\rm i}$ is equal to the pre-LBV orbital separation, and we have $M_{\rm H-env}/M_{\rm b,0} \simeq 1/3$ for Gaia BH1 and BH2, which leads to $a_{\rm i}/a_{\rm f}\simeq1/2$. After the LBV phase, the orbit keeps widening due to the WR winds, while the eccentricity obtained during the LBV phase is preserved. For isotropic mass loss \citep[Eq.\,16.20 in ][]{Tauris2006}, we find that the
WR winds widen the orbit by a factor of two, i.e., $a_{\rm f} \simeq a_{\rm obs}/2$, where $a_{\rm obs}$ denotes the currently observed semi-major axis. Therefore, the pre-LBV orbital separation is about 1/4th of the observed semi-major axis, resulting in $80\rso$ and $280\rso$ for the progenitors of Gaia BH1 and Gaia BH2. The sizes of LBVs range from tens to hundreds of solar radii. E.g., AG Carinae exhibits a radius changing from  $\sim80\rso$ to above $160\rso$ \citep{Groh2011ApJ...736...46G}. The estimated orbital separations before and after the LBV stage of our models are well compatible to these numbers.

It is beyond our scope to produce best-fit progenitor models for the Gaia BH systems.
However, the example case explored above shows that our scenario of LBV-driven mass ejections may offer a viable explanation for the origin of Gaia BH1 and Gaia BH2.

\begin{table}[t]
    \centering
    \caption{Properties of Gaia black hole systems}
    \begin{tabular}{l|ccc}\hline\hline
   System & Gaia BH1 & Gaia BH2 & Gaia BH3 \\\hline
    $\porb$ [d] &185.6& 1277&4234 (11.6\,yr)\\
        Eccentricity&0.45&0.52&0.73\\
    BH mass [$\mso$] &9.62$\pm$0.18&8.9$\pm$0.3&32.7$\pm$0.8\\
    Companion &&&\\
    - mass [$\mso$]&0.93$\pm$0.05&1.07$\pm$0.19&0.76$\pm$0.05\\
    - metallicity&solar-like& slightly low& low\\
    ~~$[$Fe/H$]$ & -0.2 & -0.22 & -2.56 \\
    \hline
    Reference&(1) \& (2)&(3)&(4)\\
    \hline
    \end{tabular}
    \tablefoot{(1) \citet{El-Badry2023MNRAS.518.1057E}; (2) \citet{Chakrabarti2023AJ....166....6C}; (3) \citet{El-Badry2023MNRAS.521.4323E}; (4) \citet{Gaia2024A&A...686L...2G}.}
    \label{tab:GaiaBH}
\end{table}

\subsection{Trigger mechanisms of mass ejection at periastron\label{sec:dis-trigger}}

Observationally, LBVs display two types of eruptions, S~Doradus eruptions \citep{Wolf1989A&A...217...87W} and  giant eruptions, such as observed in $\eta$~Carinae \citep{Smith2003} and P~Cygni \citep{Smith2006ApJ...638.1045S}. While the mechanism of the giant eruptions remains unclear, the S~Doradus eruptions likely occur in stars that reach or exceed their Eddington-limit, which leads to a loosely-bound inflated envelope of typically $10^{-3}\mso$ \citep{Grafener2012A&A...538A..40G,Sanyal2015,Grassitelli2021A&A...647A..99G,Cheng2024ApJ...974..270C,Pauli2026}. In this situation,
small variations of the stellar wind mass-loss rate can induce strong structural changes of the inflated envelope, which may reproduce the observed S~Doradus variability \citep{Grassitelli2021A&A...647A..99G}. 
For this to occur, the mass of the stars needs to exceed $\sim30\mso$ in the Milky Way, and $\sim40\mso$ in the SMC \citep{Sanyal2017A&A...597A..71S}.

When we envision a massive primary star to undergo envelope inflation prior to Roche lobe filling, the binary orbit will likely be eccentric
because tidal circularisation is inefficient in relatively wide binaries, and steady stellar winds do not considerably alter the natal eccentricity of the binaries.
So far, the effect of envelope inflation in eccentric binaries has not been investigated. However, when the primary expands, it will encounter the situation of Roche-lobe filling at periastron passage. Since the inflated envelopes are radiation pressure dominated, they are only loosely bound to the star. Therefore, it may be plausible that such envelopes are simply kicked off when the primary moves through its periastron. If this happens, the inflated envelope will grow back and the primary finds itself in the same situation as before, except that over time the stellar mass is reduced and the Eddington factor increased, which makes the occurrence of the following mass ejections at periastron easier. 

The current binary star, $\eta$~Carinae, may give a potential counterpart of this situation \citep{Hillier2001ApJ...553..837H,Smith2018MNRAS.480.1466S,Strawn2023MNRAS.519.5882S}, which shows series of brightening episodes \citep{Smith2011,Smith2017} and mass-loss enhancements \citep{Corcoran2001ApJ...547.1034C} during periastron passages.  Whereas not the whole hydrogen-rich envelope of the LBV primary stars is lost in this way, our model likely captures the main features of such events.

\subsection{Apsidal Advance\label{sec:dis-prece}}

Apsidal advance refers to the changing of orbital orientation within the orbital plane, which can happen if the mass ejection does not occur exactly at periastron or apastron (i.e., $\vv{\upsilon}_{\rm orb}\,\cdot\,\vv{r}\neq0$ in Eq.\,\ref{eq:LRL-vector}), or if a third body exists \citep{vonZeipel1910AN....183..345V,Lidov1962P&SS....9..719L,Kozai1962AJ.....67..591K,Lei2018MNRAS.481.4602L}. In particular, a considerable fraction of massive stars are known to be in triple systems \citep{Moe2017ApJS..230...15M}. 
We expect that apsidal advance can significantly reduce the velocity of the centre of mass but does not affect the evolution of other parameters. 
A detailed consideration of apsidal advance is beyond the scope of this work. In the following, we estimate its effect by assuming that the mass ejection occurs at periastron and that the periastron rotates at a constant angular velocity ($\omega_{\rm ap}$; see Appendix\,\ref{app:precession} for details), which was taken to be 
\begin{equation}
    \omega_{\rm ap} = \frac{2\pi}{10^4 P_{\rm orb,0}}.
\end{equation}
We find that all the example models in Fig.\,\ref{fig:example-track} end up with $\upsilon_{\rm CoM}$ below $5\kms$ due to apsidal advance (Fig.\,\ref{fig:vcom-prece}). Therefore, our predicted space velocities should be treated as an upper limit.

\section{Conclusion\label{sec:conclusion}}
 
A lack of observed long-period post-mass-transfer WR star binaries compared to predictions challenges our understanding of the evolution of initially wide O star binaries and has strong implications for the formation of merging BBHs. In this work, we explored the scenario that very massive wide O star binaries experience orbital evolution that is driven by LBV-type mass ejection occurring at periastron passages. We have derived the corresponding equations, presented a case study for the eccentric WR binary, WR~140, and performed population synthesis predictions for the WR stars in the SMC.

Different from the classical mass-transfer-driven orbital evolution usually adopted in current binary evolution models, the mass-ejection-driven orbital evolution increases orbital periods, enhances eccentricities, and produces measurable space velocities. We find that with reasonable initial conditions, it can self-consistently explain the formation of WR~140-like systems. The adopted parameters for the mass ejections are chosen to adhere to our current understanding of LBVs. 

Our population synthesis calculations are based on a dense grid of detailed binary evolution models computed with the SMC metallicity \citep{Wang2020,Xu2025A&A...704A.218X} and the WR+O population derived from it \citep{Xu2025A&A...704A.218X}. 
With the empirical eccentricity function in \citet{Sana2012} for the initial eccentricities of our model binaries, our Simulation1 predicts statistically 5.3 close WR+O binaries formed through the mass-transfer-driven orbital evolution, 3.8 wide WR+O binaries formed through the mass-ejection-driven orbital evolution, and 1.9 runaway WR stars, which well reproduce the observed WR star population in the SMC, 4 close WR+O binaries and 7 apparently single WR stars, including 2 runaways \citep{Schootemeijer2024}. 
The mass-ejection WR+O binaries would be difficult to identify observationally, due to their high eccentricities (0.4--1) and long orbital periods (peaking at $10^3\,$d). Hence, our scenario leads to WR binaries that are harder to identify as such in radial velocity studies. Notably, based on their increased orbital separations, they would be easier to find by GAIA. Given the uncertainties in the initial eccentricity function, the apparent absence of long-period WR+O binaries can also be explained by the binary disruption during the LBV stage if initially wide O star binaries are more eccentric than we assumed.
Future detections of more WR~140-like systems, particularly in the Magellanic Clouds, will provide further insights into the evolution of initially wide O star binaries.

Whereas not explored in detail, we find that our scenario is also capable of providing progenitor evolution models for the systems Gaia BH1 and Gaia BH2, which consist of $\sim10\mso$ BHs with solar type stars in wide eccentric orbits (Sect.\,\ref{gaiabh}). It has implications for the formation of merging BBHs through the CEE channel (Sect.~\ref{sec:merging-bbh}). If a WR star formed by mass-ejection-driven orbital evolution can produce a BH, our model predicts the existence of long-period and eccentric BH+O binaries. Such systems may experience further mass transfer in this eccentric configuration \citep{Sepinsky2007,Sepinsky2009ApJ...702.1387S,Rocha_2025,Parkosidis2025arXiv250905243P,Parkosidis2025arXiv251107190P} or even a second phase of mass-ejection-driven orbital evolution. Merging BBHs can hardly be formed in either case. The CEE channel remains possible only if tides in long-period and eccentric BH+O binaries are strong enough to circularise the orbits.

\begin{acknowledgements} 
The authors thank the referee for the constructive feedback on the manuscript.
X.-T.X.\ thanks Dong Lai and Abel Schootemeijer for helpful discussions.
X.-T.X.\ was supported by the Tsung-Dao Lee postdoctoral fellowship at the Tsung-Dao Lee Institute (TDLI).
This research was supported in part by grant NSF PHY-2309135 to the Kavli Institute for Theoretical Physics (KITP), the National Natural Science Foundation
of China (No. 12525304), and the Strategic Priority Research Program of the Chinese
Academy of Sciences (No. XDB1160201).
X.-D.L.\ was supported by the National Key Research and Development Program of China (2021YFA0718500), the Natural Science Foundation of China under grant No.\ 12041301 and 12121003.
A.H.\ was supported by the Australian Research Council (DP240101786, DP240103174) and by the Alexander von Humboldt Foundation.

\end{acknowledgements}

\bibliographystyle{aa} 
\bibliography{Xu_SMC}

\appendix

\onecolumn

\section{Tidal circularisation\label{app:tides}}

With the increasing of the Roche-lobe filling factor at periastron, tidal circularisation becomes more and more efficient. In our calculation, we have ignored the tidal circularisation during the mass-ejection-driven orbital evolution. In this section, we estimate the reduction in eccentricity related to tides and compare with the eccentricity pumping caused by mass ejection.

\def\dd{\text{d\,}}

The inflated envelope of a star near the Eddington limit is convective and contains a mass of about $10^{-3}\mso$ \citep[][; cf., Sect.\,\ref{sec:dis-trigger}]{Sanyal2015}. We hence estimated the tidal circularisation by using the equations for convective damping \citep{Hut1981,Hurley2002}, which are
\begin{equation}
    \frac{\dd e}{\dd t} = -27 \left(\frac{k}{T}\right)q(1+q) \left(\frac{R}{a}\right)^8\frac{e}{(1-e^2)^{13/2}}\left[f_3(e^2) - \frac{11}{18}(1-e^2)^{3/2}f_4(e^2)\frac{\Omega_{\rm spin}}{\Omega_{\rm orb}}\right]
\end{equation}
where $e$ is the eccentricity, $q$ is the mass ratio of the binary (secondary/primary), $R$ is the radius of the primary star, $a$ is the semi-major axis, $\Omega_{\rm spin}$ is the rotational angular velocity of the star, $\Omega_{\rm orb}$ is the mean angular velocity of the binary, the $f_3(e^2)$ and $f_4(e^2)$ functions are  
\begin{equation}
    f_3(e^2) = 1+\frac{15}{4}e^2 + \frac{15}{8}e^4 + \frac{5}{64}e^6,
\end{equation}
and
\begin{equation}
    f_4(e^2) = 1+\frac{3}{2}e^2 + \frac{1}{8}e^4,
\end{equation}
and the $k/T$ term is
\begin{equation}
    \frac{k}{T} = \frac{2}{21} \frac{f_{\rm conv}}{\tau_{\rm conv}}\frac{M_{\rm env}}{M}\,\text{yr}^{-1}, 
\end{equation}
where $M_{\rm env}$ is the mass of the inflated envelope, $M$ is the stellar mass, and $\tau_{\rm conv}$ is the convection turnover time, which depends of the size of the envelope ($R_{\rm env}$) and stellar luminosity ($L$), i.e.,
\begin{equation}
    \tau_{\rm conv} = 0.4311\left[\frac{M_{\rm env} R_{\rm env} (R - 1/2R_{\rm env})}{3L}\right]^{1/3} \,\text{yr}^{-1}.
\end{equation}
For a crude estimation, we took $\Omega_{\rm spin} = \Omega_{\rm orb}$, which gives $f_{\rm conv} = 1$, and  we used the Roche-lobe radius given by the fitting formula in \citet{Eggleton1983} for the radius of the primary star with $R_{\rm env} = 1/2R$.

The luminosity at terminal-age main sequence of $30\mso$ is about $3\times10^{5}\lso$ according to the models in \citet{Xu2025A&A...704A.218X}. Considering a $24\mso$ companion star with an orbital period of 40\,d and an eccentricity of 0.4, the reduction in eccentricity over one orbital period is
\begin{equation}
    (\Delta e)_{\rm tide} = \left(\frac{\dd e}{\dd t}\right)\times P_{\rm orb}\sim 10^{-7}.
    \label{eq:tides}
\end{equation}

Meanwhile, the eccentricity pumping caused by a mass ejection at periastron is given by
\begin{equation}
    (\Delta e)_{\rm LBV} = e_1 - e_0 =\frac{\Delta M + e_0 M_{\rm b,0} - e_0 M_{\rm b,0} - e_0 \Delta M}{M_{\rm b,0} - \Delta M}=\frac{\Delta M }{M_{\rm b,0} - \Delta M}(1- e_0). 
\end{equation}
Taking $\Delta M = 10^{-3}\mso$, $M_{\rm b,0}=54\mso$ (i.e., $30\mso+24\mso$), and $e_0=0.4$, we obtained 
\begin{equation}
    (\Delta e)_{\rm LBV} \sim 10^{-5},
\end{equation}
which is much more significant than the reduction caused by tides ($10^{-7}$; Eq.\,\ref{eq:tides}). Since the ejected mass can be much higher than $10^{-3}\mso$, we conclude that tidal circulation can be safely ignored during the mass-ejection-driven orbital evolution.

\section{Mass ejection at apastron\label{app:apastron_case}}

In the main text, we focus on the mass ejection at periastron. In this section, we extend the calculation to apastron for completeness. By replacing $r_{\rm per,0}$ by the separation at apastron, $r_{\rm ap,0}$, and $\upsilon_{\rm per,0}$ by the relative velocity at apastron, 
\begin{equation}
    \upsilon_{\rm ap,0} = \sqrt{\frac{GM_{\rm b,0}}{a_0}\frac{1-e_0}{1+e_0}} = \left(\frac{GM_{\rm b,0}}{P_{\rm orb,0}}2\pi\right)^{1/3}\sqrt{\frac{1-e_0}{1+e_0}},
    \label{eq:v-ap0}
\end{equation}
in the equations in Sect.\,\eqref{sec:me-drive-model}, we can obtain the orbital parameters after the $n$th apastron passage, which are
\begin{equation}
    a_n=a_0 (1+e_0) \frac{M_{\rm b,0}-n\Delta M}{M_{\rm b,0}(1+e_0)-2 n\Delta M},
\end{equation}
\begin{equation}
e_n^2=\left(\frac{e_0 M_{\rm b,0}-n\Delta M}{M_{\rm b,0} - n\Delta M}\right)^2,
\end{equation}
\begin{equation}
    P_{\text{orb},n}^2 =P_{\rm orb,0}^2\frac{(1-n\Delta M/M_{\rm b,0})^2(1+e_0)^3}{[(1+e_0)-2n\Delta M / M_{\rm b,0}]^3},
\end{equation}
\begin{equation}
    \upsilon_{\text{CoM},n} = \frac{n\Delta M M_{\rm s2}}{M_{\rm b,0} (M_{\rm b,0}-n\Delta M)}\upsilon_{\rm ap,0}.
\end{equation}
Different from the periastron case, eccentricities have two solutions depending on ejected mass, which are
\begin{equation}
e_n=
\begin{cases}
    (e_0 M_{\rm b,0}-n\Delta M)/(M_{\rm b,0} - n\Delta M) {\rm ~~for~}n\Delta M\leq e_0 M_{\rm b,0}\\
    (n\Delta M-e_0 M_{\rm b,0})/(M_{\rm b,0} - n\Delta M) {\rm ~~for~}n\Delta M> e_0 M_{\rm b,0}
\end{cases}.
\end{equation}
These two solutions lead to the following relations,
\begin{itemize}
    \item $n\Delta M < e_0 M_{\rm b,0}$: $a_n(1+e_n) = a_0(1+e_0)$, which means that the separation at apastron of the post-mass-ejection orbit is equal to that of the pre-mass-ejection orbit;
    \item $n\Delta M > e_0 M_{\rm b,0}$: $a_n(1-e_n) = a_0(1+e_0)$, which means that the separation at periastron of the post-mass-ejection orbit is equal to that at apastron of the pre-mass-ejection orbit.
\end{itemize}
The orientation of the post-mass-ejection orbit also changes differently in both cases.
The Laplace-Runge-Lenz vector before the mass ejection is 
\begin{equation}
    \vv{e}_0 =\left[\frac{\upsilon_{\rm ap,0}^2}{GM_{\rm b,0}} - \frac{1}{r_{\rm ap,0}}\right]\vv{r}_{\rm ap,0}
    =\frac{-e_0}{a_0(1+e_0)}\vv{r}_{\rm ap,0},
\end{equation}
and, after ejecting $n\Delta M$ mass, it becomes
\begin{equation}
    \vv{e}_1 =\left[\frac{\upsilon_{\rm ap,0}^2}{G(M_{\rm b,0}-n\Delta M)} - \frac{1}{r_{\rm ap,0}}\right]\vv{r}_{\rm ap,0}.
\end{equation}
Taking $n\Delta M=e_0M_{\rm b,0}$ leads to
\begin{equation}
    \frac{\upsilon_{\rm ap,0}^2}{G(M_{\rm b,0}-eM_{\rm b,0})} - \frac{1}{r_{\rm ap,0}} = 0.
\end{equation}
Therefore, 
\begin{itemize}
    \item $n\Delta M < e_0 M_{\rm b,0}$: $\vv{e}_0 \cdot \vv{e}_1 = |\vv{e}_0 \cdot \vv{e}_1 |$, which means that the orientation of apastron of the pre-mass-ejection orbit is the same as for the post-mass-ejection orbit;
    \item $n\Delta M > e_0 M_{\rm b,0}$: $\vv{e}_0 \cdot \vv{e}_1 = -|\vv{e}_0 \cdot \vv{e}_1 |$, which means that the orientation of apastron of the pre-mass-ejection orbit becomes the orientation of periastron of the post-mass-ejection orbit.
\end{itemize}

\section{Establishing the formulae in Sect.\,\ref{sec:long-term-evo} using mathematical induction\label{app:math-induction}}

\textbf{Theorem.}
After the $n$th mass ejection at periastron, the orbital parameters are given by
\begin{equation}
    a_n=a_0 (1-e_0) \frac{M_{\rm b,0}-n\Delta M}{M_{\rm b,0}(1-e_0)-2 n\Delta M},
    \label{eq:appB-1}
\end{equation}
\begin{equation}
    e_n=\frac{n\Delta M+e_0 M_{\rm b,0}}{M_{\rm b,0} - n\Delta M},
    \label{eq:appB-2}
\end{equation}
\begin{equation}
    P_{\text{orb},n}^2 =P_{\rm orb,0}^2\frac{(1-n\Delta M/M_{\rm b,0})^2(1-e_0)^3}{[(1-e_0)-2n\Delta M / M_{\rm b,0}]^3},
    \label{eq:appB-3}
\end{equation}
\begin{equation}
    \upsilon_{\text{CoM},n} =\frac{n\Delta M M_{\rm s2}}{M_{\rm b,0} (M_{\rm b,0}-n\Delta M)}\upsilon_{\rm per,0}.
    \label{eq:appB-4}
\end{equation}

\textbf{Proof.} We establish the formulae by induction on $n$.
\begin{itemize}
    \item \textbf{Base case.} Taking $n=1$, Eqs.\,\eqref{eq:appB-1}-\eqref{eq:appB-4} are consistent with our result for one periastron passage (Sect.\,\ref{sec:post-ME-orbit}).

    \item \textbf{Inductive hypothesis.} We assume that the identity of Eqs.\,\eqref{eq:appB-1}-\eqref{eq:appB-4} holds for a given number of periastron passage, $k$, which is
    \begin{equation}
    a_k=a_0 (1-e_0) \frac{M_{\rm b,0}-k\Delta M}{M_{\rm b,0}(1-e_0)-2 k\Delta M},
\end{equation}
\begin{equation}
    e_k=\frac{k\Delta M+e_0 M_{\rm b,0}}{M_{\rm b,0} - k\Delta M},
\end{equation}
\begin{equation}
    P_{\text{orb},k}^2 =P_{\rm orb,0}^2\frac{(1-k\Delta M/M_{\rm b,0})^2(1-e_0)^3}{[(1-e_0)-2k\Delta M / M_{\rm b,0}]^3},
\end{equation}
\begin{equation}
    \upsilon_{\text{CoM},k} =\frac{k\Delta M M_{\rm s2}}{M_{\rm b,0} (M_{\rm b,0}-k\Delta M)}\upsilon_{\rm per,0}.
\end{equation}

    \item \textbf{Inductive step.} In the following, we show that this identity holds for $n=k+1$. After the $k$th periastron passage, the total mass of the binary becomes $(M_{\rm b,0} - k\Delta M)$, and the mass ejected during the $(k+1)$th periastron passage is $\Delta M$ according to our assumption. 
    The orbital parameters after the $(k+1)$th periastron passage, $(a_{k+1},\,e_{k+1}, P_{{\rm orb},k+1},\,\upsilon_{{\rm CoM},k+1})$, can be obtained by using our result for one mass ejection, Eqs.\,\eqref{aper1}-\eqref{Pper1} and \eqref{vcom1}. By replacing the pre-mass-ejection binary mass, $M_{\rm b,0}$, by $(M_{\rm b,0}-k\Delta M)$, we obtain 
    \begin{equation}
    \begin{split}
    a_{k+1}&=a_k (1-e_k) \frac{(M_{\rm b,0}-k\Delta M)-\Delta M}{(M_{\rm b,0}-k\Delta M)(1-e_k)-2 \Delta M}
    =a_0 (1-e_0) \frac{[M_{\rm b,0}-(k+1)\Delta M]}{M_{\rm b,0}(1-e_0)-2(k+1) \Delta M},
    \end{split}
\end{equation}
\begin{equation}
\begin{split}
    e_{k+1}&=\frac{\Delta M+e_k (M_{\rm b,0}-k\Delta M)}{(M_{\rm b,0}-k\Delta M) -\Delta M}
    =\frac{(k+1)\Delta M+e_0 M_{\rm b,0}}{M_{\rm b,0} -(k+1)\Delta M},
    \end{split}
\end{equation}
\begin{equation}
\begin{split}
    P_{\text{orb},k+1}^2 &=P_{{\rm orb},k}^2\frac{[1-\Delta M/(M_{\rm b,0}-k\Delta M)]^2(1-e_k)^3}{[(1-e_k)-2\Delta M / (M_{\rm b,0}-k\Delta M)]^3}
    =P_{\rm orb,0}^2\frac{[1-(k+1)\Delta M/M_{\rm b,0}]^2(1-e_0)^3}{[(1-e_0)-2(k+1)\Delta M / M_{\rm b,0}]^3},
    \end{split}
\end{equation}    
\begin{equation}
\begin{split}
    \upsilon_{\text{CoM},k+1} &=\upsilon_{{\rm CoM},k}+\frac{\Delta M M_{\rm s2}}{(M_{\rm b,0} - k\Delta M)[(M_{\rm b,0}-\Delta M)-k\Delta M]}\upsilon_{\rm per,0}
    =\frac{(k+1)\Delta M M_{\rm s2}}{M_{\rm b,0} [M_{\rm b,0}-(k+1)\Delta M]}\upsilon_{\rm per,0}.
\end{split}
\end{equation}
These equations are consistent with the expectation of Eqs.\,\eqref{eq:appB-1}-\eqref{eq:appB-4} by taking $n=k+1$.

\item \textbf{Conclusion.} we conclude that Eqs.\,\eqref{eq:appB-1}-\eqref{eq:appB-4} hold for all positive integer $n$ if the binary stays bound (i.e., $e_{n}<1$).
\end{itemize}

\section{WR~140: MESA fitting method\label{app:mesa-fit}}

We scanned through the comprehensive binary evolution model grid of \citet{Jin2025} to search for the MESA binary models matching the observed properties of WR~140 and its companion. We required the models to satisfy the following three constraints simultaneously:
\begin{itemize}
    \item the system undergoes Case B mass transfer;
    \item after mass transfer, the primary becomes a stripped star (WR star) with a mass of 10.31$\pm0.45\,\mso$ during its core helium burning phase, and it would appear as a WC star with a surface carbon mass fraction above 0.2;
    \item the secondary is an O-type star with a mass of 29.27$\pm1.14\mso$.
\end{itemize}
Since the MESA models were computed assuming stable mass transfer in circular orbits, we did not impose any constraints on the orbital periods and eccentricities of the models.
We find that models with an initial primary mass of about $40\mso$, an initial mass ratio close to 1 (i.e., initial secondary masses of 30--35$\mso$), and initial orbital periods in a range of about 200--300\,d could satisfy all the three constraints simultaneously. These binary models would appear as WC+O binaries for a few 0.1 Myr, which is a significant fraction of the core helium burning timescale of the primary stars.

\section{Predictions with fixed initial eccentricities\label{app:popsync-fixed-e0-more}}

\begin{figure*}[ht!]
        \centering
        \includegraphics[width=0.43\linewidth]{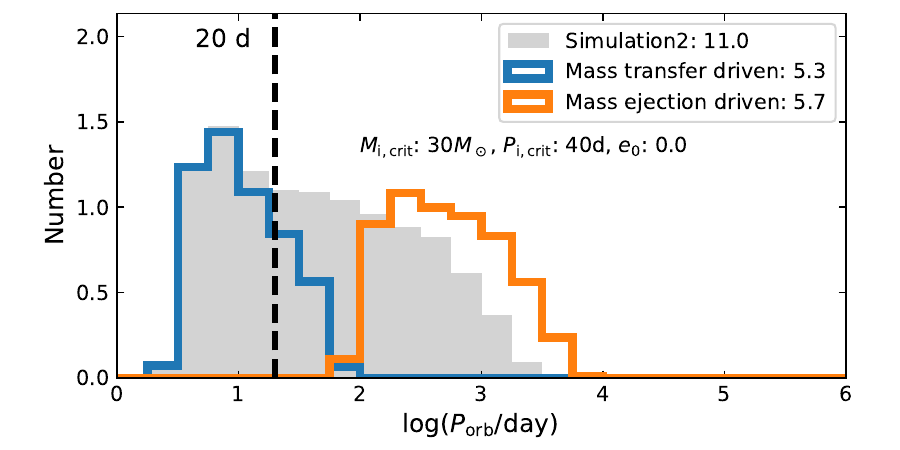}
        \includegraphics[width=0.43\linewidth]{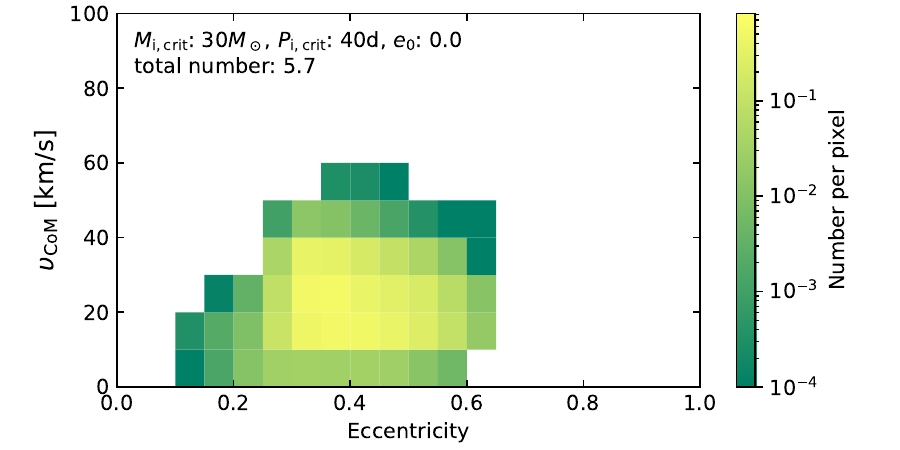}
        \includegraphics[width=0.43\linewidth]{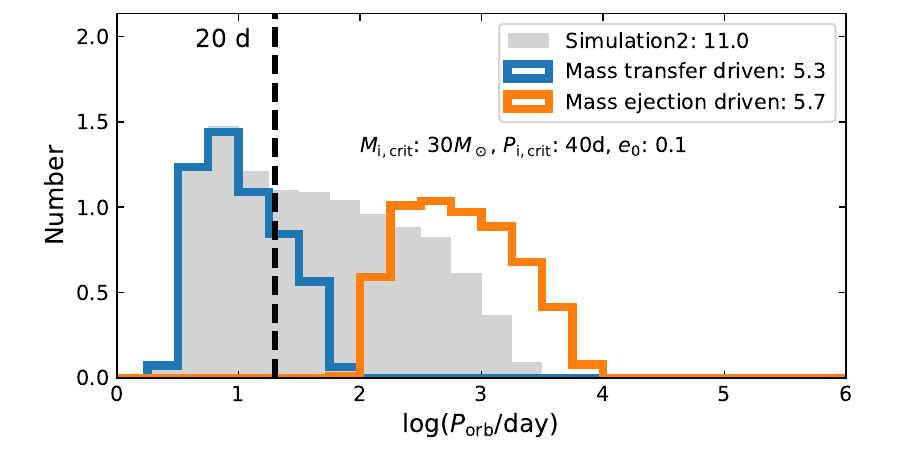}
        \includegraphics[width=0.43\linewidth]{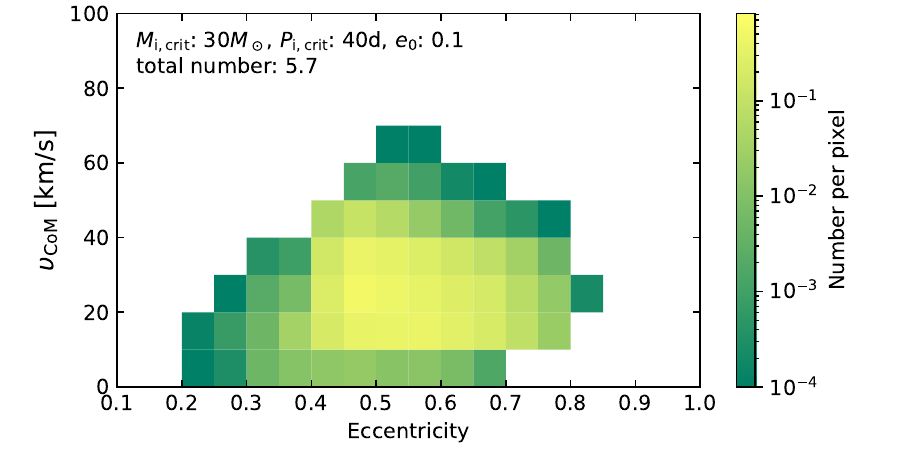}
        \includegraphics[width=0.43\linewidth]{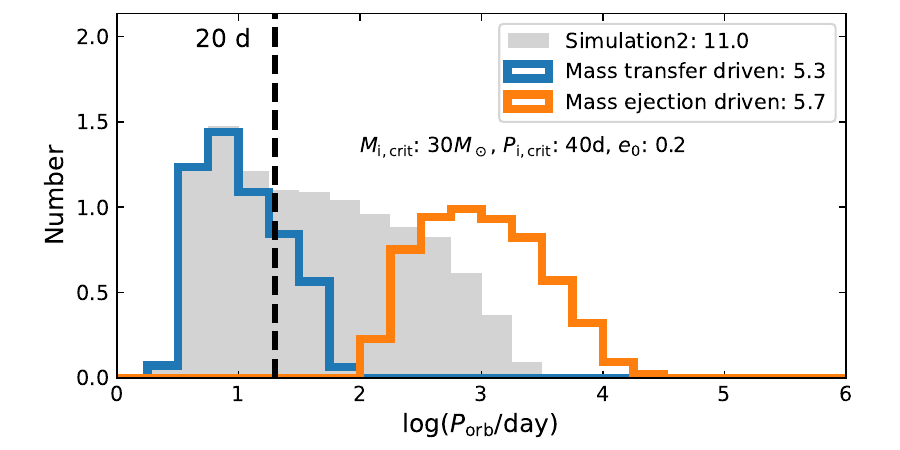}
        \includegraphics[width=0.43\linewidth]{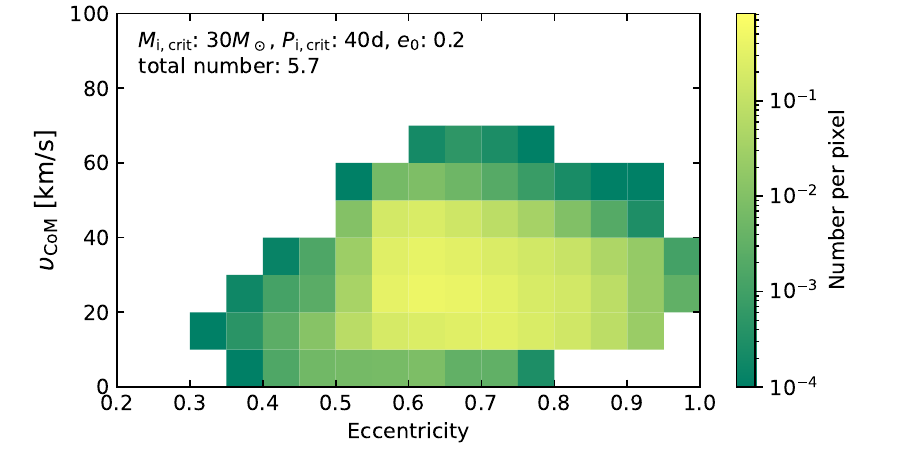}
        \includegraphics[width=0.43\linewidth]{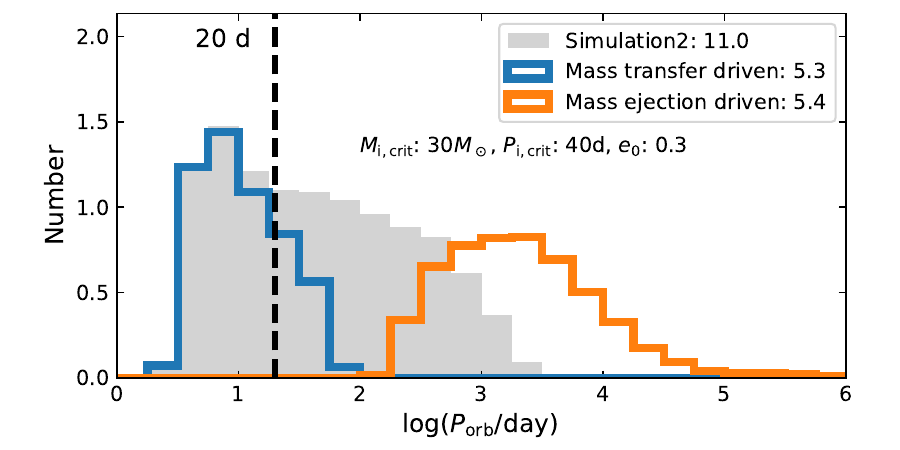}
        \includegraphics[width=0.43\linewidth]{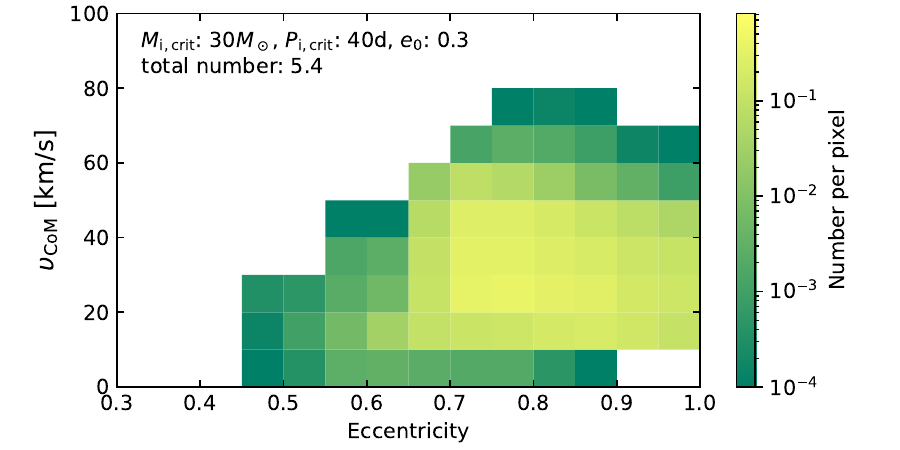}
        \includegraphics[width=0.43\linewidth]{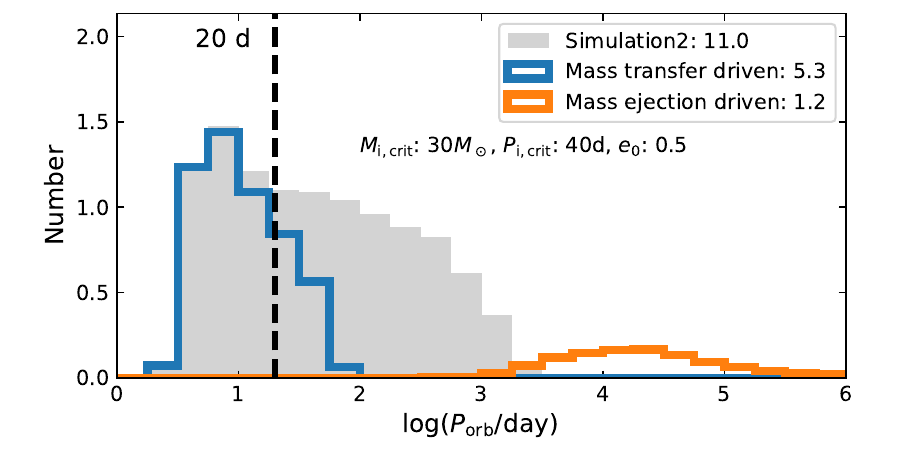}
        \includegraphics[width=0.43\linewidth]{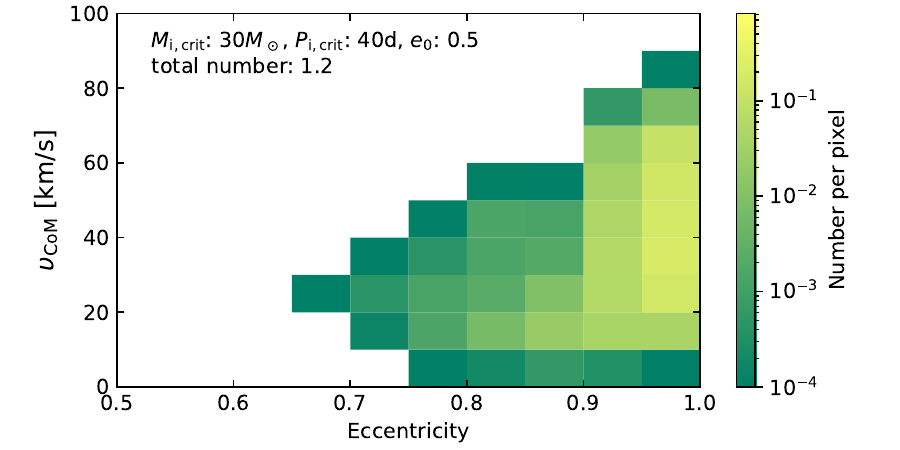}
        \caption{Population synthesis predictions with fixed initial eccentricities, which are 0.0, 0.1, 0.2, 0.3, and 0.5, from top to bottom. The left column shows the predicted orbital period distributions of the mass-transfer WR+O binaries (blue) and mass-ejection WR+O binaries (orange), where the predicted numbers are indicated in the text. The vertical lines correspond to the observed longest orbital period of the SMC WR+O binaries. The grey background is the prediction of Simulation2, which is the fiducial model in \citet{Xu2025A&A...704A.218X} but with a higher star formation rate. The right column shows the predicted population of the mass-ejection WR+O binaries on the eccentricity-$\upsilon_{\rm CoM}$ plane, where the predicted number in each pixel is colour-coded.}
        \label{fig:popsync-fixed-e0-more}
\end{figure*}

In this section, we present the population synthesis predictions with fixed initial eccentricities, 0.0, 0.1, 0.2, 0.3, and 0.5 (Fig.\,\ref{fig:popsync-fixed-e0-more}). Consistent with the fiducial population synthesis model, $M_\text{crit}$ and $P_\text{crit}$ are taken to be $30\mso$ and $40\,$d. Predictions with initial eccentricities above 0.6 are not presented, since all the systems entering the mass-ejection-driven orbital evolution are  disrupted. 

With the increasing of initial eccentricities, the orbital periods of the mass-ejection WR+O binaries are shifted towards the long-period regime. Binary disruption starts to take place with $e_0=0.3$. Correspondingly, the expected eccentricities also increase from $\sim0.5$ to near 1, while the velocities of the centre of mass stay at around 20--40$\kms$.

\section{Predictions by different population synthesis models\label{app:param-study}}

\begin{figure}[ht!]
    \centering
    \includegraphics[width=0.43\linewidth]{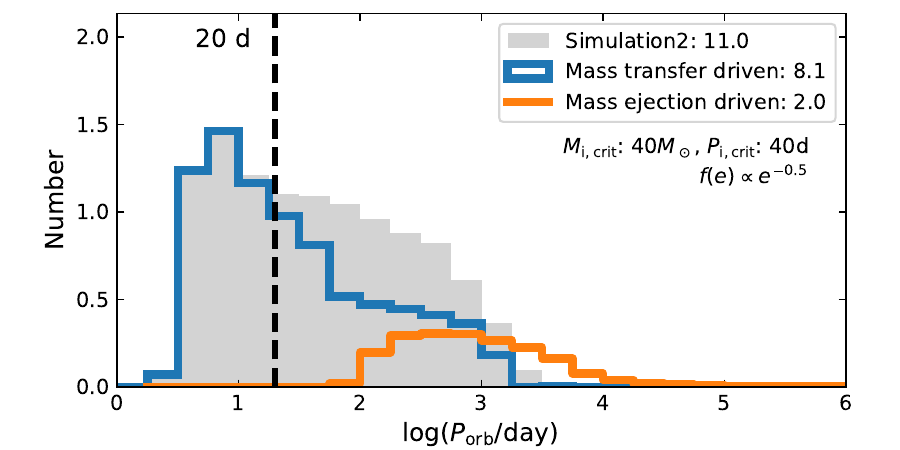}
    \includegraphics[width=0.43\linewidth]{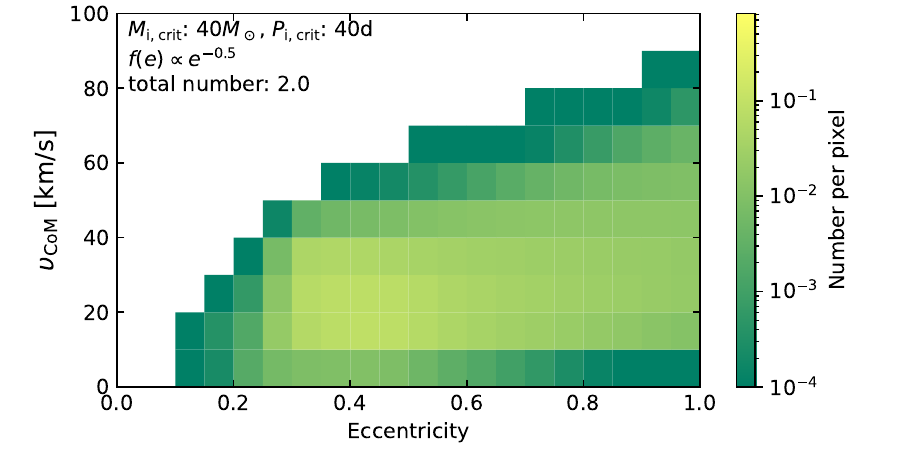}
    \includegraphics[width=0.43\linewidth]{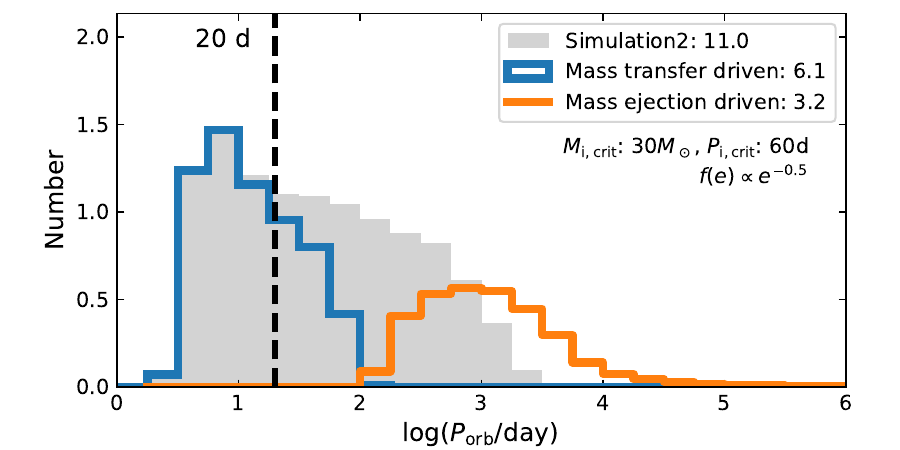}
    \includegraphics[width=0.43\linewidth]{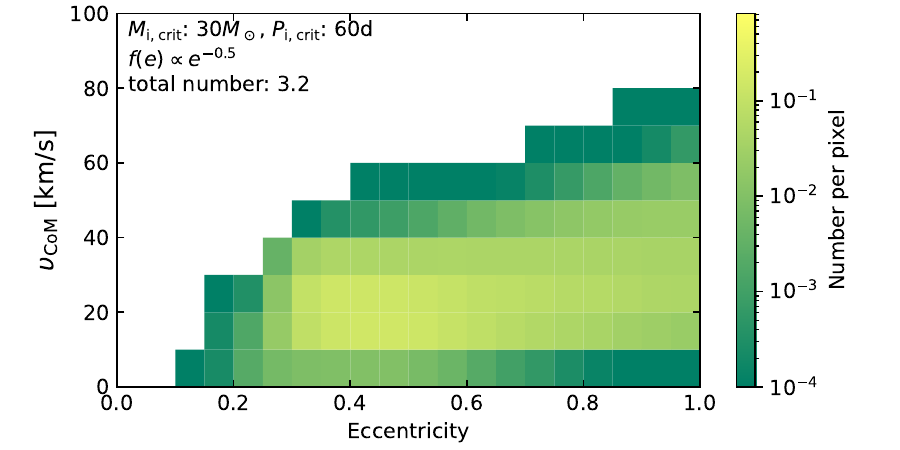}
    \includegraphics[width=0.43\linewidth]{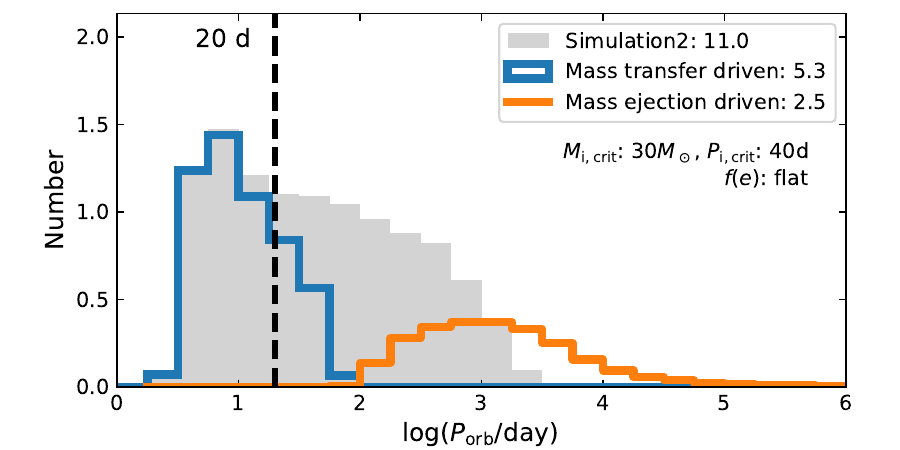}
    \includegraphics[width=0.43\linewidth]{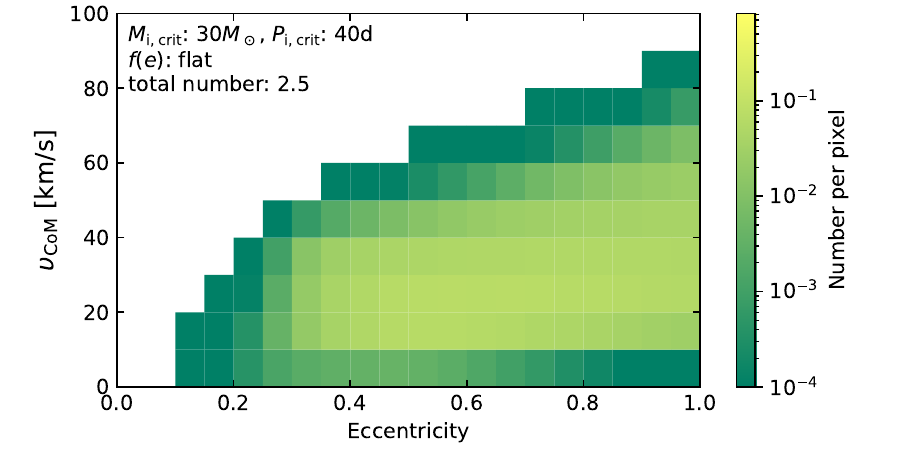}
    \caption{Predicted populations by different different population synthesis models. The colours and legends have the same meaning as in Fig.\,\ref{fig:popsync-fixed-e0-more}. From top to bottom rows, we present the predictions in the High-$M_{\rm crit}$ model ($M_\text{crit}=40\mso$), High-$P_\text{crit}$ model ($P_{\rm crit}=60\,$d), and Flat-$f_e$ model ($\eta=0$).}
    \label{fig:popsync-other-models}
\end{figure}

In this section, we present the distribution functions derived from the High-$M_{\rm crit}$, High-$P_\text{crit}$, and Flat-$f_e$ models (see Tab.\,\ref{tab:popsync-model} for the definitions of the models). The predicted populations are presented in Fig.\,\ref{fig:popsync-other-models}. We refer to Sect.\,\ref{sec:param-study} for detailed descriptions.

\section{Apsidal advance\label{app:precession}}

In the main text, we have ignored the effects of apsidal advance, which can significantly reduce the velocity of the centre of mass by changing the angle between $\vv{\upsilon}_{\rm per,0}$ and $\vv{\upsilon}_{\rm CoM,0}$ in Eq.\,\eqref{eq:vcom1}. In this section, we estimate the effect of apsidal advance by assuming that the periastron rotates at a fixed angular momentum velocity, $\omega_{\rm ap}$.

\begin{figure}[ht!]
    \sidecaption
    \includegraphics[trim=20mm 0mm 10mm 35mm, clip,width=0.3\linewidth]{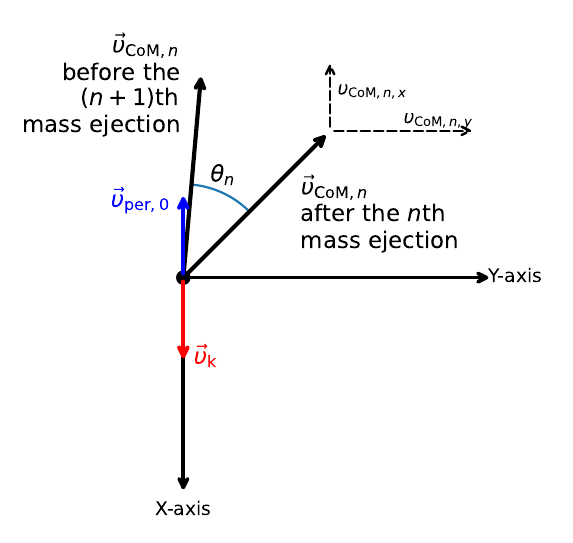}
    \caption{Coordinate system adopted for estimating the effect of apsidal advance on the velocity of centre of mass. Due to apsidal advance, the orientation of the velocity of the centre of mass, $\vv{\upsilon}_{\text{CoM},n}$, changes by an angle of $\theta_{n}$ from the $n$th to $(n+1)$th mass ejection. The momentum kick induced by the $(n+1)$th mass ejection, $\vv{\upsilon}_{\rm k}$, has the opposite direction as the relative velocity at periastron, $\vv{\upsilon}_{\rm per,0}$.}
    \label{fig:prece-coord}
\end{figure}

Considering a coordinate system corotating with the orbit, the X-axis has the opposite direction as the relative velocity at periastron, and the Y-axis points from the companion star to the mass-losing star (Fig.\,\ref{fig:prece-coord}). After the $n$th periastron passage, the velocity of the centre of mass and orbital period are defined as $\vv{\upsilon}_{\rm CoM,n}$ and $P_{\rm orb,n}$. We denote the components on the X- and Y-axis of $\vv{\upsilon}_{\rm CoM,n}$ as ${\upsilon}_{\rm CoM,n,x}$ and ${\upsilon}_{\rm CoM,n,y}$, respectively. In the following, we derive the velocity of the centre of mass, $\upsilon_{\rm CoM,n+1}$, after the $(n+1)$th periastron passage by taking into account apsidal advance.  Due to apsidal advance, the orientation of $\vv{\upsilon}_{\rm CoM,n}$ changes an angle of $\theta_{n}$, which is
\begin{equation}
    \theta_{n} = -\omega_{\rm ap}\,P_{\text{orb},n},
\end{equation}
where the negative sign is caused by the choice of the corotating coordinate system. For example, if the orientation of the orbit rotates clockwise for a faraway observer, $\vv{\upsilon}_{\rm CoM,n}$ should rotate anti-clockwise in this corotating coordinate system. Then, at the $(n+1)$th mass ejection, $\vv{\upsilon}_{\rm CoM,n}$ becomes
\begin{equation}
    \vv{\upsilon}_{\text{CoM},n}=
    \begin{bmatrix}
    \upsilon_{\text{CoM},n,x}\cos \theta_n -\upsilon_{\text{CoM},n,y}\sin\theta_n\\
    \upsilon_{\text{CoM},n,x}\sin \theta_n +\upsilon_{\text{CoM},n,y}\cos\theta_n
    \end{bmatrix}.
    \label{eq:vcom-n}
\end{equation}
Let $\vv{\upsilon}_{\rm K}$ be the kick velocity induced by the $(n+1)$th mass ejection, and the velocity of the centre of mass after the $(n+1)$th mass ejection be given by 
\begin{equation}
    \vv{\upsilon}_{\text{CoM},n+1} = \vv{\upsilon}_{\text{CoM},n} +\vv{\upsilon}_{\rm K}.
    \label{eq:vcom-n+1}
\end{equation}
As we have shown in Sect.\,\ref{sec:post-ME-orbit}, the direction of $\vv{\upsilon}_{\rm K}$ is the opposite direction of the relative velocity at periastron, which is $(1,\,0)$ in the considered coordinate system. Hence $\vv{\upsilon}_{\rm K}$ is given by
\begin{equation}
    \vv{\upsilon}_{\rm K} = \frac{\Delta M M_{\rm s2}}{(M_{\rm b,0}-n\Delta M) [(M_{\rm b,0}-n\Delta M)-\Delta M]}\upsilon_{\rm per,0}\,
\begin{bmatrix}
1  \\
0
\end{bmatrix},
\label{eq:vk}
\end{equation}
where $(M_{\rm b,0}-n\Delta M)$ represents the total mass of the binary after the $n$th mass ejection.
Combining Eqs.\,\eqref{eq:vcom-n}, \eqref{eq:vcom-n+1}, and \eqref{eq:vk}, we obtained the X- and Y-axis components, $\upsilon_{\text{CoM},n+1,x}$ and $\upsilon_{\text{CoM},n+1,y}$, of $\vv{\upsilon}_{\text{CoM},n+1}$ as the following
\begin{equation}
\begin{split}
    \upsilon_{\text{CoM},n+1,\text{x}}&=\bigg|(\upsilon_{\text{CoM},n,x}\cos \theta_n -\upsilon_{\text{CoM},n,y}\sin\theta_n) 
    + \frac{\Delta M M_{\rm s2}}{(M_{\rm b,0}-n\Delta M) [(M_{\rm b,0}-n\Delta M)-\Delta M]}\upsilon_0\bigg|,
\end{split}
\end{equation}
and
\begin{equation}
    \upsilon_{\text{CoM},n+1,\text{y}}=\bigg|(\upsilon_{\text{CoM},n,x}\sin \theta_n +\upsilon_{\text{CoM},n,y}\cos\theta_n)\bigg|.
\end{equation}
Then, we have 
\begin{equation}
    (\upsilon_{{\rm CoM},n+1})_\text{ap} = \sqrt{\upsilon_{\text{CoM},n+1,x}^2 + \upsilon_{\text{CoM},n+1,y}^2}.
\end{equation}

\begin{figure}[ht!]
    \centering
    \includegraphics[width=0.4\linewidth]{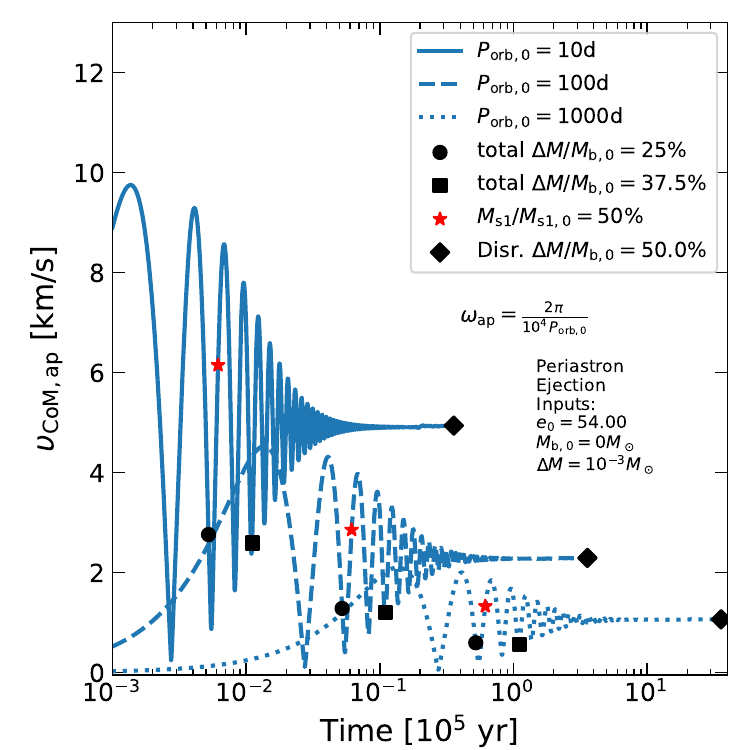}
    \includegraphics[width=0.4\linewidth]{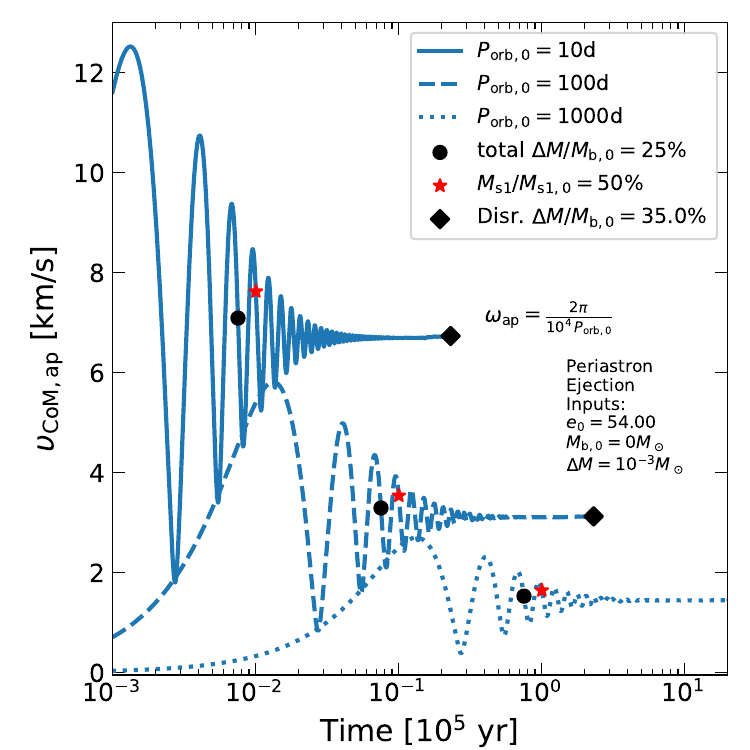}
    \caption{Effect of apsidal advance on the velocity of the centre of mass of the example models in Fig.\,\ref{fig:example-track}.  The apsidal advance angular velocity was taken to be $\omega_{\rm ap}=2\pi \,(10^{4}P_{\rm orb,0})^{-1}$. The two panels correspond to two initial eccentricities, 0.0 (left) and 0.3 (right). 
    Each panel contains three curves, computed with different initial orbital periods, 10\,d (solid line), 100\,d (dashed line), and 1000\,d (dotted line). The markers indicate the points where the systems eject $25\%M_{\rm b,0}$ (circle), $37.5\%\mso$ (square), and $\Delta M_{\rm dis}$ (Eq.\,\ref{dm_per_disr}; diamond). The red star means that the mass-losing star ejects half of its initial mass. 
    }
    \label{fig:vcom-prece}
\end{figure}

In the following, we recalculate the velocities of the centre of mass, $\upsilon_{\rm CoM,ap}$, for the example models presented in Fig.\,\ref{fig:example-track} with the orbital orientation rotating at a constant angular velocity of 
\begin{equation}
    \omega_{\rm ap}=\frac{2\pi}{10^4 P_{\rm orb,0}}.
\end{equation}
Our result is presented in Fig.\,\ref{fig:vcom-prece}, where all models end up with $\upsilon_{\rm CoM,ap}\lesssim 5\kms$. Binaries cannot have considerable space velocities even if $\omega_{\rm ap}$ is small. Hence the predicted velocity of the centre of mass should generally be treated as an upper limit.

\end{document}